\documentclass{sig-alternate}
\usepackage{mathptmx} 
\usepackage{balance}
\usepackage{fancyhdr}
\usepackage[normalem]{ulem}
\usepackage[hyphens]{url}
\usepackage[sort,nocompress]{cite}
\usepackage[final]{microtype}
\usepackage{xspace}
\usepackage{xcolor, soul}
\usepackage{color,linegoal}
\usepackage{tabularx,ragged2e,booktabs}
\usepackage{algorithmic}
\usepackage[ruled,vlined]{algorithm2e}
\usepackage[T1]{fontenc}
\usepackage{multirow}
\usepackage{graphicx, subfigure}

\usepackage[bookmarks=true,breaklinks=true,letterpaper=true,colorlinks,linkcolor=black,citecolor=blue,urlcolor=black]{hyperref}

\newcommand{\HALCONE }{{\color{black} HALCONE \xspace}}
\newcommand{\dave}[1]{{\color{blue}\bfseries [ Dave: #1 ]}}
\newcommand{\saiful}[1]{{\color{purple}\bfseries [Saiful: #1]}}

\newcommand{\fixme}[1]{ {\color{red}\bfseries [ FIXME: #1 ]}}

\definecolor{aquamarine}{rgb}{0.5, 1.0, 0.83}
\definecolor{ashgrey}{rgb}{0.7, 0.75, 0.71}
\definecolor{atomictangerine}{rgb}{1.0, 0.6, 0.4}
\definecolor{babyblue}{rgb}{0.54, 0.81, 0.94}
\definecolor{fluorescentyellow}{rgb}{0.8, 1.0, 0.0}
\definecolor{lavender(floral)}{rgb}{0.71, 0.49, 0.86}
\definecolor{mauve}{rgb}{0.88, 0.69, 1.0}
\newcommand{\ignore}[1]{}

\DeclareRobustCommand{\Com}[1]{{\sethlcolor{aquamarine}\hl{#1}}}

\usepackage{tikz}
\newcommand*\circled[1]{\tikz[baseline=(char.base)]{
                \node[shape=circle,fill,inner sep=1pt] (char) {\textcolor{white}{#1}};}}

                \newcommand*\circld[1]{\tikz[baseline=(char.base)]{
                            \node[shape=circle,draw,inner sep=0.5pt] (char) {#1};}}

\ignore{
\newcommand{\microsubmissionnumber}{678}

\fancypagestyle{firstpage}{
  \fancyhf{}
  
  \fancyhead[C]{\vspace{15pt}\normalsize{MICRO 2020 Submission
      \textbf{\#\microsubmissionnumber} -- Confidential Draft -- Do NOT Distribute!!}}
  \fancyfoot[C]{\thepage}
}
}
\title{\HALCONE: A Hardware-Level Timestamp-based\\
Cache Coherence Scheme for Multi-GPU systems \vspace{-25pt}}
\vspace{-20pt}

\begin{document}

\author{
	Saiful~A.~Mojumder$^1$,
	Yifan~Sun$^2$,
	Leila~Delshadtehrani$^1$,
	Yenai Ma$^1$,
	Trinayan Baruah$^2$,\\
	Jos\'e~L.~Abell\'an$^3$,
	John~Kim$^4$,
	David~Kaeli$^2$,
	Ajay~Joshi$^1$\\
	\normalfont{\small $^1$ECE Department, Boston University;}
	\normalfont{\small $^2$ECE Department, Northeastern University;}\\
	\normalfont{\small $^3$CS Department, UCAM;}
	\normalfont{\small $^4$School of EE, KAIST;}\\
	\normalfont{\small \{msam, delshad, yenai joshi\}@bu.edu, \{yifansun, tbaruah, kaeli\}@ece.neu.edu,}\\
  \normalfont{\small jlabellan@ucam.edu, jjk12@kaist.edu}
  \vspace{-3mm}
  }

\maketitle

\pagestyle{plain}


\begin{abstract}

While multi-GPU (MGPU) systems are extremely popular for compute-intensive workloads, several inefficiencies in the memory hierarchy and data movement result in a waste of GPU resources and difficulties in programming MGPU systems. First, due to the lack of hardware-level coherence, the MGPU programming model requires the programmer to replicate and repeatedly transfer data between the GPUs’ memory. This leads to inefficient use of precious GPU memory. Second, to maintain coherency across an MGPU system, transferring data using  low-bandwidth and high-latency off-chip links leads to degradation in system performance. Third, since the programmer needs to manually maintain data coherence, the programming of an MGPU system to maximize its throughput is extremely challenging. To address the above issues, we propose a novel lightweight timestamp-based coherence protocol, \HALCONE, for MGPU systems and modify the memory hierarchy of the GPUs to support physically shared memory. \HALCONE replaces the Compute Unit (CU) level logical time counters with cache level logical time counters to reduce coherence traffic. Furthermore, \HALCONE introduces a novel timestamp storage unit (TSU) with no additional performance overhead in the main memory to perform coherence actions. Our proposed \HALCONE protocol maintains the data coherence in the memory hierarchy of the MGPU with minimal performance overhead (less than 1\%). 
Using a set of standard MGPU benchmarks, we observe that a 4-GPU MGPU system with shared memory and \HALCONE performs, on average, 4.6$\times$ and 3$\times$ better than a 4-GPU MGPU system with existing RDMA and with the recently proposed HMG coherence protocol, respectively. 
We demonstrate the scalability of \HALCONE using different GPU counts (2, 4, 8, and 16) and different CU counts (32, 48, and 64 CUs per GPU) for 11 standard benchmarks. Broadly, \HALCONE scales well with both GPU count and CU count. Furthermore, we stress test our \HALCONE protocol using a custom synthetic benchmark suite to evaluate its impact on the overall performance. When running our synthetic benchmark suite, the \HALCONE protocol slows down the execution time by only 16.8\% in the worst case.
\end{abstract}

\vspace{-9pt}
\section{Introduction}
\label{sec:introduction}
Multi-GPU (MGPU) systems have become an integral part of cloud services such as Amazon Web Services (AWS)~\cite{cloud2011amazon}, Microsoft Azure~\cite{copeland2015microsoft}, and Google Cloud~\cite{googlecloud}. In particular, many deep learning frameworks running on these cloud services provide support for MGPU execution to significantly accelerate the long and compute-intensive process of training deep neural networks. 
For instance, Goyal et al.~\cite{goyal2017accurate} trained ResNet-50 in only 1 hour using 256 GPUs, which would have otherwise taken more than a week using a single GPU. MGPU systems are also commonly used for parallelizing irregular graph applications~\cite{che2013pannotia,xu2014graph, shi2019realtime, wang2019excavating} and facilitating large-scale simulations in different domains including physics~\cite{zhu2018employing}, computational algebra~\cite{yamazaki2014tridiagonalization}, surface metrology~\cite{zhang2018fuzzy} and medicine~\cite{hagan2018multi}.

\begin{figure}[t] \centering
  \includegraphics[width=0.9\columnwidth]{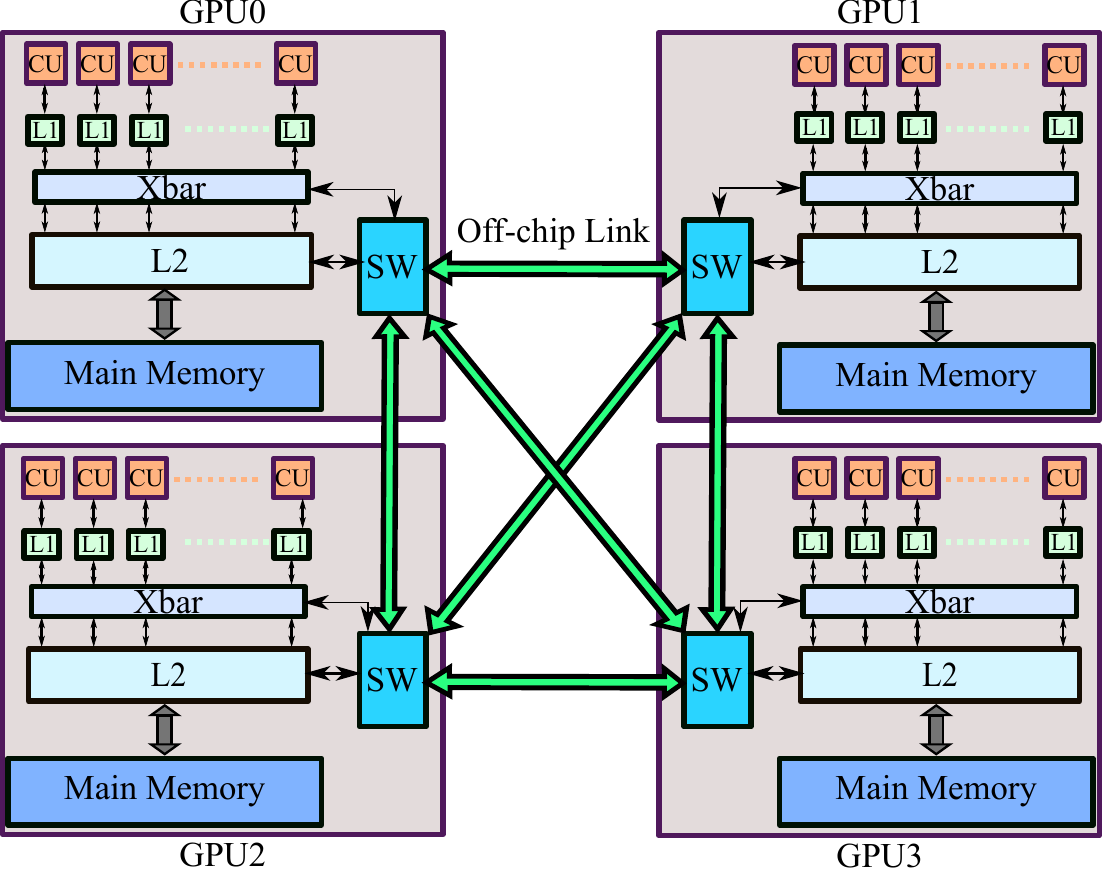} \vspace{-10pt}
  \caption{Conventional MGPU system. Switch (SW) handles the remote access requests from one GPU to another GPU. PCIe or NVLink is used as the off-chip link.}
  \vspace{-0.28in}
  \label{fig:nvlink}
\end{figure}

\sloppy{GPU applications are evolving to support ever-larger datasets and demand data communications not only within a GPU but also across multiple GPUs in the system.  As a result, the underlying programming model is evolving, and major vendors such as NVIDIA and AMD have introduced software abstractions such as shared and global memory spaces that enable sharing of data among different threads within a GPU. Furthermore, GPU-to-GPU Remote Direct Memory Access (RDMA)~\cite{unifiedmemory} was introduced so that a GPU can transfer data directly to/from another GPU, without involving the CPU, through off-chip links such as NVLink or PCIe as shown in Figure~\ref{fig:nvlink}. However, despite continuous advances in inter-GPU networking technologies, the off-chip link bandwidth is roughly 3-10$\times$ lower than local main memory (MM) bandwidth. Thus, data accesses to/from remote GPU memories, which are essential for MGPU applications, can lead to severe performance degradation due to the {\em NUMA effect}~\cite{milic2017beyond,mojumder2018profiling}.}

\begin{figure}[t] \centering
\includegraphics[width=0.8\columnwidth]{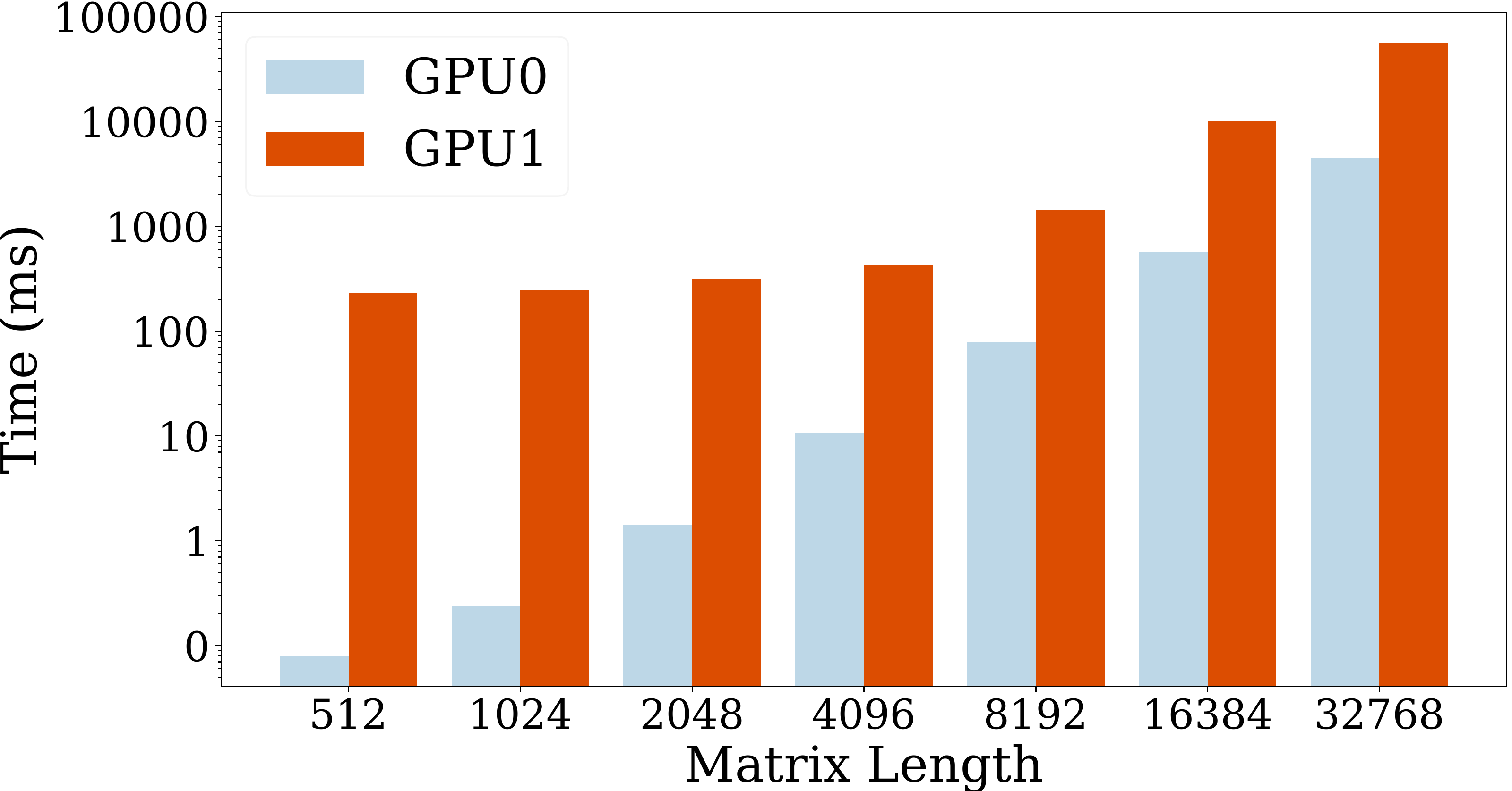} \vspace{-10pt}
\caption{Kernel time in GPU$_0$ (local) and GPU$_1$ (remote) on Volta-based NVIDIA
DGX-1 system.} \vspace{-0.26in}
\label{fig:dgx-1} \end{figure}

To highlight the large impact on performance when relying on state-of-the-art RDMA technology for GPU-to-GPU communication, we perform an experiment with a \texttt{SGEMM} kernel from the cuBLAS library~\cite{nvidia2008cublas} that is executed on an NVIDIA DGX-1~\cite{harris2017nvidia} MGPU system. More specifically, we aim to compare the impact of RDMA on kernel execution time considering two Volta GPUs (i.e., GPU$_0$ and GPU$_1$), which are connected by NVLinks that support 50 GB/s per direction. We first place the matrices in GPU$_0$'s memory and execute \texttt{SGEMM}. Then, we enable \textit{P2P direct access} to leverage RDMA and run the same \texttt{SGEMM} kernel on GPU$_1$ while the matrices reside in GPU$_0$'s memory (we refer to this as remote). Figure~\ref{fig:dgx-1} shows the results of the experiment with different matrix sizes. The kernel using local memory is $12.4\times$ (for a matrix size of $32768 \times 32768$) to $2895\times$ (for a matrix size of $512 \times 512$) faster than the kernel using remote memory. These results clearly show that RDMA is expensive, and motivate the pressing need for an alternate path to reduce the significant cost of remote accesses in MGPU systems.

Programming a GPU system also has several challenges. In particular, a program written to run on a single GPU cannot be easily ported to run on multiple GPUs.  The programmer must be aware of any data sharing, both within a single GPU and across multiple GPUs. As there is no hardware support for cache coherency in current MGPU systems, the programmer must explicitly maintain coherency while developing a GPU program. This requires  manually transferring duplicated data when needed to different GPUs' memory, using explicit barriers if previously read data is modified, or in the worst case, using atomic operations. All of these approaches require significant effort by the programmer. Moreover, contemporary GPUs only support weak data-race-free (DRF) memory consistency~\cite{hechtman2013exploring,singh2015efficiently}, so the programmer has to ensure there are no data races during program execution.

\ignore{
In summary, today's DNN and graph applications have inherent sharing that presently is managed by the programmer while using a single GPU because of lack of hardware coherency support. On top of that the size of the networks and graphs has exhausted the compute and memory resources of a single GPU, so multi-GPU execution is a necessity and for that coherency is needed across GPUs.
}

To leverage the true capability of MGPU systems and ease MGPU programming, we need efficient hardware-level inter-GPU coherency. Extending well-known directory-based or snooping-based CPU coherence protocols such as MESI, MOESI, etc. is not suitable for an MGPU system~\cite{Ziabari2016Taco}. This is because the MGPU system with its thousands of parallel threads per GPU produces a much larger number of simultaneous memory requests than CPUs, which translates into a prohibitively high degree of coherence traffic in these traditional protocols~\cite{singh2013cache}. A promising solution to alleviate coherence traffic is to use self-invalidation by relying on temporal coherence~\cite{singh2013cache}. Previous timestamp based solutions targeting a single GPU used global time (the TC protocol)~\cite{singh2013cache} or logical time (the G-TSC protocol)~\cite{tabbakh2018g} to maintain coherency across the L1\$s and L2\$s of a GPU. However, none of the previous timestamp based work has addressed coherency issues in an MGPU system. 

To provide support for efficient intra-GPU and inter-GPU coherency, in this paper we propose \HALCONE - a new timestamp-based hardware-level cache coherence scheme for tightly-coupled MGPU systems with physically shared main memory (named MGPU-SM for short)\footnote{An MGPU system can be designed by cobbling together a collection of MGPU nodes, and leveraging a message passing layer, such as MPI~\cite{manian2019} to work collectively on a single application.  An alternate approach is to consider an MGPU system that is more tightly-coupled and uses shared memory (i.e., MGPU-SM)~\cite{dgx1}.  The former involves more programmer effort to support MGPU execution, while the latter relies on the hardware and runtime system to support a version of shared memory. In this paper, we focus on improving the performance and scalability of an MGPU-SM system.}. \HALCONE provides hardware-level coherency support for inter- and intra-GPU data sharing using a single-writer-multiple-reader (SWMR) invariant. \HALCONE introduces a new cache-level logical time counter to reduce Core-to-Cache traffic\footnote{Request traffic is reduced by up to 41.7\% and the response traffic is reduced by up to 3.1\% in the memory hierarchy.} and a novel timestamp storage unit (TSU) to keep track of cache blocks' timestamps. We strategically place the TSU outside the critical path of memory requests, as it is accessed in parallel with the MM, thereby avoiding any performance overhead. 
Although \HALCONE is inspired by G-TSC~\cite{tabbakh2018g},  G-TSC is not suitable as is for both NUMA- and UMA-based MGPU systems due to the large area required for storing timestamps and the high performance overhead due to additional traffic to maintain coherency. Prior MGPU-based solutions such as CARVE~\cite{young2018combining} and HMG~\cite{hmg} minimize RDMA overhead by introducing a simple VI cache coherence protocol. However, the scalability of these solutions is limited because of the large amount of coherency traffic required to transfer via off-chip links with low bandwidth and high latency. Moreover, HMG relies on a scoped memory model which increases the programmer's burden~\cite{sinclair2015efficient}. \HALCONE is a holistic highly-scalable solution that reduces coherence traffic and eases programming of MGPU systems. The main takeaways of our work are: \break
\ignore{
There are MGPU communication libraries, such as
NCCL~\cite{nvidiadeveloper_2018}, that provide MPI-like collectives for MGPU
data transfer. However, the NCCL library also avoids data races by breaking down
data into chunks and pipelining data transfers using either memcpy or RDMA
direct accesses. To use NCCL, a collective such as AllReduce creates N sets of
data if there are N GPUs. Most of the time, replicating data is unnecessary as
that data will not be used by all the GPUs. For instance, the data parallel
model used for DNN training workloads requires aggregation of the gradients for
weight updates using an AllReduce operation. This process replicates the
aggregated gradients in all the GPUs, though only a single GPU uses the
gradients to update weights. As a result, we waste a lot of GPU memory.
Moreover, it can lead to a large number of remote GPU accesses, which impacts
the application's performance~\cite{mojumder2018profiling}.}
\ignore{
\jose{In the paper, we rely on sequential consistency. However, As in
G-TSC, the \HALCONE protocol could rely on release consistency that needs programming
effort. So your previous sentence might no be true.} \dave{I also worry about
claiming a lot about consistency. The paper is more about coherency versus
supporting weak consistency.}}
\textbf{Design:} We are the first to propose a fully coherent MGPU system with physically shared MM. Our MGPU-SM system eliminates NUMA accesses to MM, as well as the need to transfer data back and forth between GPUs. We propose a novel timestamp-based inter-GPU and intra-GPU coherence protocol called \HALCONE to ensure seamless data sharing in an MGPU system. \HALCONE leverages the concept of {\em logical timestamp}~\cite{plakal1998lamport} and reduces the overall traffic by replacing a compute unit (CU)-level timestamp counter at the cache level. In MGPU-SM, \HALCONE uses a new timestamp storage unit (TSU) that operates in parallel with MM to avoid any performance overhead.\\
\textbf{Evaluation:} We evaluate our MGPU-SM system with \HALCONE using MGPUSim~\cite{sun2019mgpusim}. For evaluation, we use standard benchmarks and custom synthetic benchmarks (that enforce read-write data sharing) to stress test the \HALCONE protocol. Compared to an MGPU system with RDMA, our MGPU-SM system with \HALCONE performs, on average, 4.6$\times$ better across the standard benchmarks. Our \HALCONE protocol adds minimal performance overhead (1\% on average) to the standard benchmarks. Compared to the most recently proposed MGPU system with an HMG coherence protocol~\cite{hmg}, \HALCONE performs, on average, 3$\times$ better. Stress tests performed using our synthetic benchmark suite show that \HALCONE can result in up to 16.8\% performance degradation for extreme cases, but it lowers programmer burden.  We also show that an MGPU system with \HALCONE scales well with GPU count and CU count per GPU.
\vspace{-10pt}
\section{Background}
\vspace{-5pt}
\label{sec:background}

\vspace{-2pt}
\subsection{Communication and Data Sharing}
\label{sec:Existing MGPU System}
To share data across GPUs in MGPU systems, a variety of communication mechanisms are available. A GPU can use \textit{P2P memcpy} to copy data to and from the memory of a remote GPU. This \textit{P2P memcpy} approach essentially replicates the data in different memory modules~\cite{Ziabari2016Taco}. To avoid this data replication, \textit{P2P direct access} (called RDMA in this paper) can be used where a GPU can access data from a remote GPU memory without copying it to its MM. However, \textit{P2P direct access} leads to high latency remote data accesses, which causes performance degradation~\cite{young2018combining}. \ignore{ that was also discussed in Section~\ref{sec:introduction}.}

Current NVIDIA GPUs provide a page fault mechanism in a virtually unified address space called unified memory (UM)\footnote{UM provides the abstraction of a virtually unified memory combining the CPU's and GPUs' physical memories.}. Under UM, the data is initially allocated on the CPU. When a GPU tries to access this data, it triggers a page fault and the GPU driver serves the page fault by fetching the required page from another device. However, this page fault mechanism is known to introduce serialization in accessing pages and can hurt GPU performance~\cite{kim2020batch,griffin}.

\ignore{

To make programming simpler, NVIDIA has introduced the unified memory (UM)~\cite{NvidiaCuda} model. UM simplifies MGPU programming by creating a virtual unified memory combining the CPUs' and GPUs' physical memories. Under UM, the data is initially allocated on the CPU. When a GPU tries to access this data, it triggers a page fault and the GPU driver serves the page fault.\ignore{ The GPU driver and runtime then handle this page fault and migrate the data to the faulting GPU's memory. Future accesses to a page located on one GPU from another GPU can either be handled by migrating the page again to the requesting GPU or using \textit{P2P direct access}.} Using page migration to service page fault requests is known to significantly slowdown the execution of a program~\cite{pearson2019evaluating, zheng2016towards, agarwal2015unlocking}. \ignore{ UM also provides an API to allow GPUs to use CPU memory directly using Zerocopy~\cite{NvidiaCuda}. In this way, UM also aims at addressing GPU memory capacity issues for applications that demand large amount of memory. However, the additional memory space is available at the expense of slow data transfer~\cite{otterness2017evaluation}.}In general, none of these methods help us harness the full potential of a MGPU system.
}
\vspace{-5pt}
\subsection{Timestamp Based Coherency}
\label{sec:G-TSC}

In this section, we briefly explain the operation of G-TSC protocol~\cite{tabbakh2018g}, which was proposed to maintain coherency in a single GPU. G-TSC protocol assigns a logical timestamp (\texttt{warpts}) to each CUs in the GPU. Table~\ref{tab:terms} provides definitions for the terminology used to describe the G-TSC and also our proposed \HALCONE protocol.\\

\vspace{-0.15in}
\textbf{Read Operation:} A read request from a CU contains the \texttt{warpts} and the address. Each block in the L1\$ \footnote{Throughout this paper, we use L1\$ to refer to L1\$ vector cache unless specified otherwise. } has a read timestamp (\texttt{rts}) and write timestamp (\texttt{wts}). If the block is present and the \texttt{warpts} falls within the range between \texttt{wts} and \texttt{rts} (i.e. the \texttt{lease}), it is a cache hit. Otherwise, the read request is treated as an L1\$ miss and is forwarded to the L2\$. This read request to L2\$ contains the address, the \texttt{wts} of the block and the \texttt{warpts}. If the \texttt{wts} value for the request is set to 0, it means a compulsory miss occurred at L1\$, and L2\$ must respond with the data and the timestamps. A non-zero value for \texttt{wts} implies the block exists in L1\$, but the timestamp expired.

Upon receiving a read request, the L2\$  checks if the block exists in the L2\$. If it does not exist, L2\$ sends a read request to the MM. If the \texttt{warpts} of the request from L1\$ is within the \texttt{lease} for that block in the L2\$, the L2\$  also compares the \texttt{wts} from the L1\$ request and the \texttt{wts} of the block in the L2\$. If both \texttt{wts} values are the same, it means that the data was not modified by another CU and simply that the lease expired in the L1\$. In that case, the L2\$ extends the lease by increasing the \texttt{rts} and sends the new \texttt{rts} and \texttt{wts} values to the L1\$. If the \texttt{wts} values do not match, it implies that the data was modified by a different CU. Hence, L2\$ sends both data and new timestamps to the L1\$ . \\

\vspace{-0.15in}
\textbf{Write Operation:} Write requests are handled using a write-though policy from L1\$ to L2\$, and L1\$ adopts a no-write-allocate policy. To result in a write hit in a cache, the \texttt{warpts} needs to be  within the \texttt{lease} of the requested cache block. Otherwise, it is considered a write miss. When there is a write hit in the L1\$, the data is written in the L1\$, but the access to the data is locked until L2\$ is updated and L2\$ sends the updated timestamps for the data. It is necessary to lock access to the block to ensure that the \texttt{warpts} is updated correctly using the \texttt{wts} value that the L1\$  receives from the L2\$. Any discrepancy in updating \texttt{warpts} may result in an incorrect ordering of memory access requests. If there is an L1\$ write miss, the data is directly sent to the L2\$ to complete the write operation.  For an L2\$ write miss, the L2\$ sends a write request to the MM. If we get a write hit at L2\$, the L2\$ writes the data to the block in the L2\$ and updates the timestamps for that block. L2\$ then sends the updated \texttt{rts} and \texttt{wts} values to the L1\$ .

On both read and write operations, the responses from L1\$ to CU
contain the \texttt{wts} value from the most recent memory operation. Based on
this \texttt{wts} value, CU updates its \texttt{warpts}.

\begin{table}[t]
\small
\caption{Terminologies and definitions}
\label{tab:terms}
\begin{tabular}{p{0.4in}|p{2.6in}}
\toprule
Term & Definition\\
\midrule \texttt{physical time}  & The wall clock time of an operation.\\
\texttt{logical time}  & The logical counter maintained by a
component (e.g., CU and cache). \\
\texttt{warpts}  & The current logical time of a CU. In the G-TSC protocol, the memory operations are ordered based on the \texttt{warpts}.  \\
\texttt{cts}   & The current
logical times of a cache. Each cache has a \texttt{cts} that is updated based on the last memory operation.  \\
\texttt{block} & An entry containing address, data, and associated timestamps in the caches. \\
\texttt{wts}   &  The write timestamp of a cache block. It represents the
logical time when a write operation is visible to the processors \\
\texttt{rts}   & The read timestamp of a cache block. It represents
the logical time until which reading the cache block is valid \\
\texttt{lease} & The difference between \texttt{rts} and \texttt{wts}.
The data in the cache block is valid only if \texttt{cts} or \texttt{warpts} is within the lease. \\
\texttt{RdLease}   & Lease assigned to a block after a read operation is executed.\\
\texttt{WrLease}   & Lease assigned to a block after a write operation is executed.\\
\texttt{memts} & The memory time stamp that represents the logical read timestamp
that the memory assigns to a cache block. \\
\bottomrule
\end{tabular}
\vspace{-15pt}
\end{table}

\vspace{-10pt}
\section{\HALCONE in MGPU-SM Systems}
\label{sec:tsm}
\begin{figure*}[!htb]
  \centering
  \begin{minipage}{.25\textwidth}
    \centering
    \includegraphics[width=\linewidth]{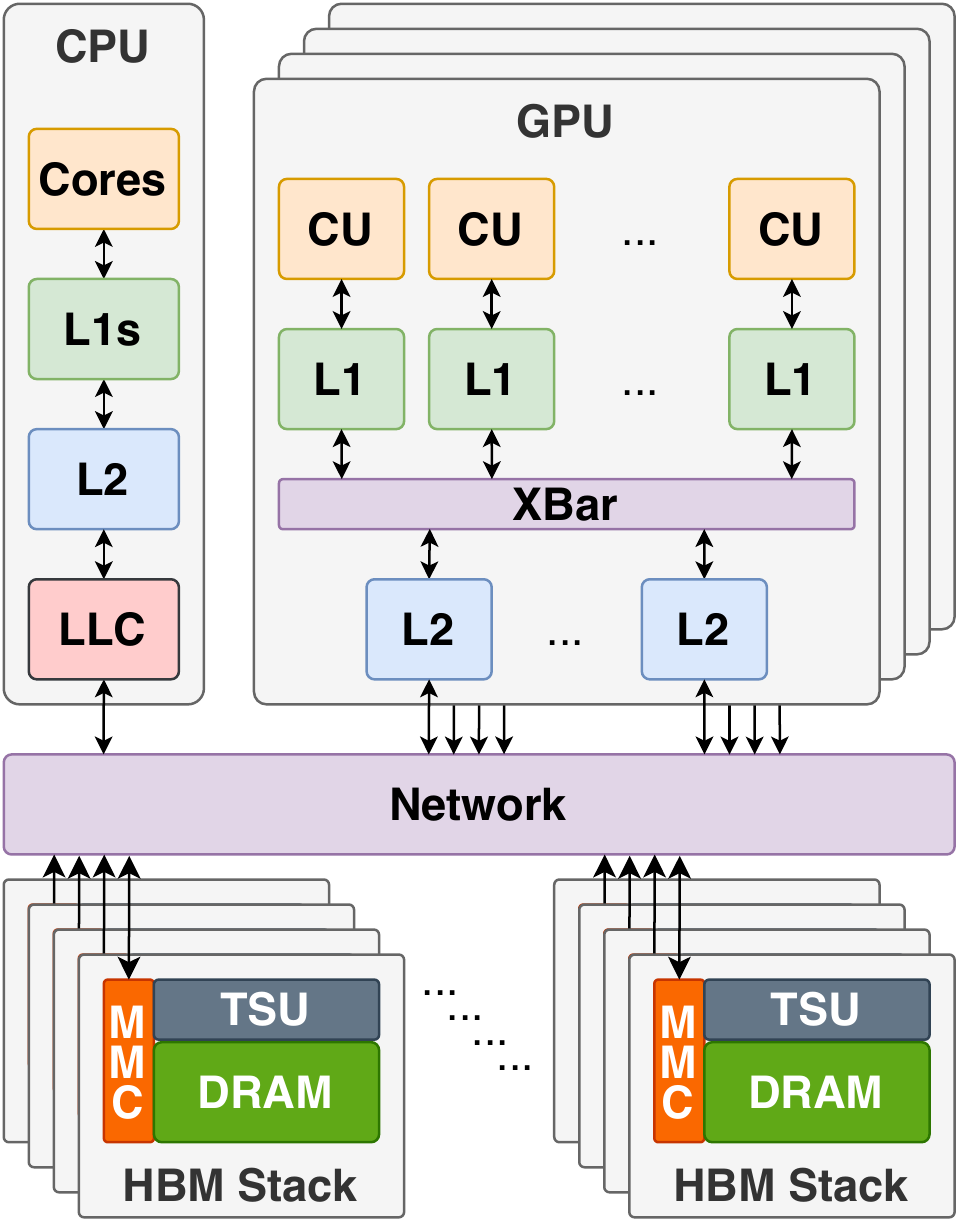} \vspace{-15pt}
    \caption{MGPU with shared main memory.} \vspace{-10pt}
    \label{fig:tsm1}
  \end{minipage}%
  \hspace{0.05\textwidth}
  \begin{minipage}{0.65\textwidth}
    \centering
    \includegraphics[width=\linewidth]{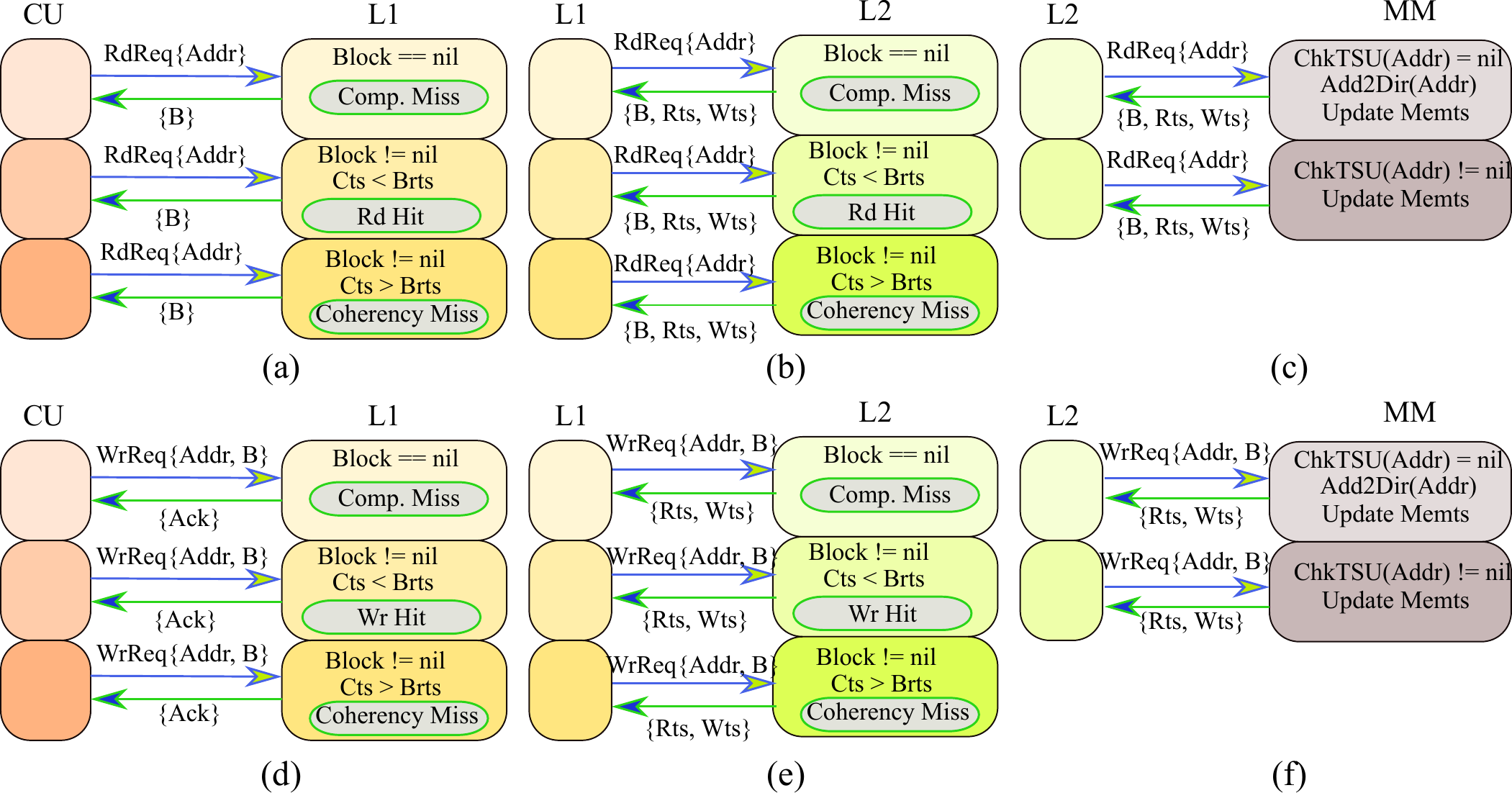} \vspace{-22pt}
    \caption{Transactions between (a) a CU and an L1\$ for read operations, (b) an L1\$ and an L2\$ for read operations, (c) an L2\$ and the MM  for read operations, (d) a CU and an L1\$ for write operations, (e) an L1\$ and an L2\$ for write operations, and (f) an L2\$ and the MM for write operations.}
    \label{fig:coh}
  \end{minipage}
  \vspace{-14pt}
\end{figure*}
We present the working fundamentals of our proposed \HALCONE using an example MGPU system with 4 GPUs that share the MM. HALCONE builds on top of the G-TSC protocol~\cite{tabbakh2018g}, which was designed for intra-GPU coherence but cannot be readily applicable to MGPU systems. Maintaining coherence across multiple GPUs is more challenging as the L1\$s of a GPU can only interact with their own L2\$. We need to maintain coherence across multiple L2\$s across different GPUs. A straightforward extension of G-TSC would be to add timestamps to each block in the MM leading to significant area overhead as we would need space to store the timestamps of each block of data in the MM. G-TSC also needs to maintain a logical time counter at the compute unit level, which needs to send timestamps i.e. \texttt{warpts} back and forth between CUs and L1\$s leading to additional traffic overhead. As we show later in the paper (Section~\ref{subsec:halconeprotocol}), to maintain coherence in MGPU systems we do not need to maintain a logical time counter in the CU level, but instead, each L1\$ and L2\$ can maintain the individual counters. This helps reduce request traffic. In addition, while G-TSC simply provides the same lease for both reads and writes, we provide different lease values for reads and writes. By doing so, as we will see in Section~\ref{sec:ts}, this benefits the exploitation of temporal locality. To elaborate, each write operation moves the logical time counter ahead by the write lease value. If we use the same lease for read as for write, each write operation will lead to self-invalidation of the previously read block.

\vspace{-3pt}
\subsection{An MGPU System with Shared Memory}
\ignore{
\begin{figure*}[!htb]
  \centering
  \begin{minipage}{.3\textwidth}
    \centering
    \includegraphics[width=\linewidth]{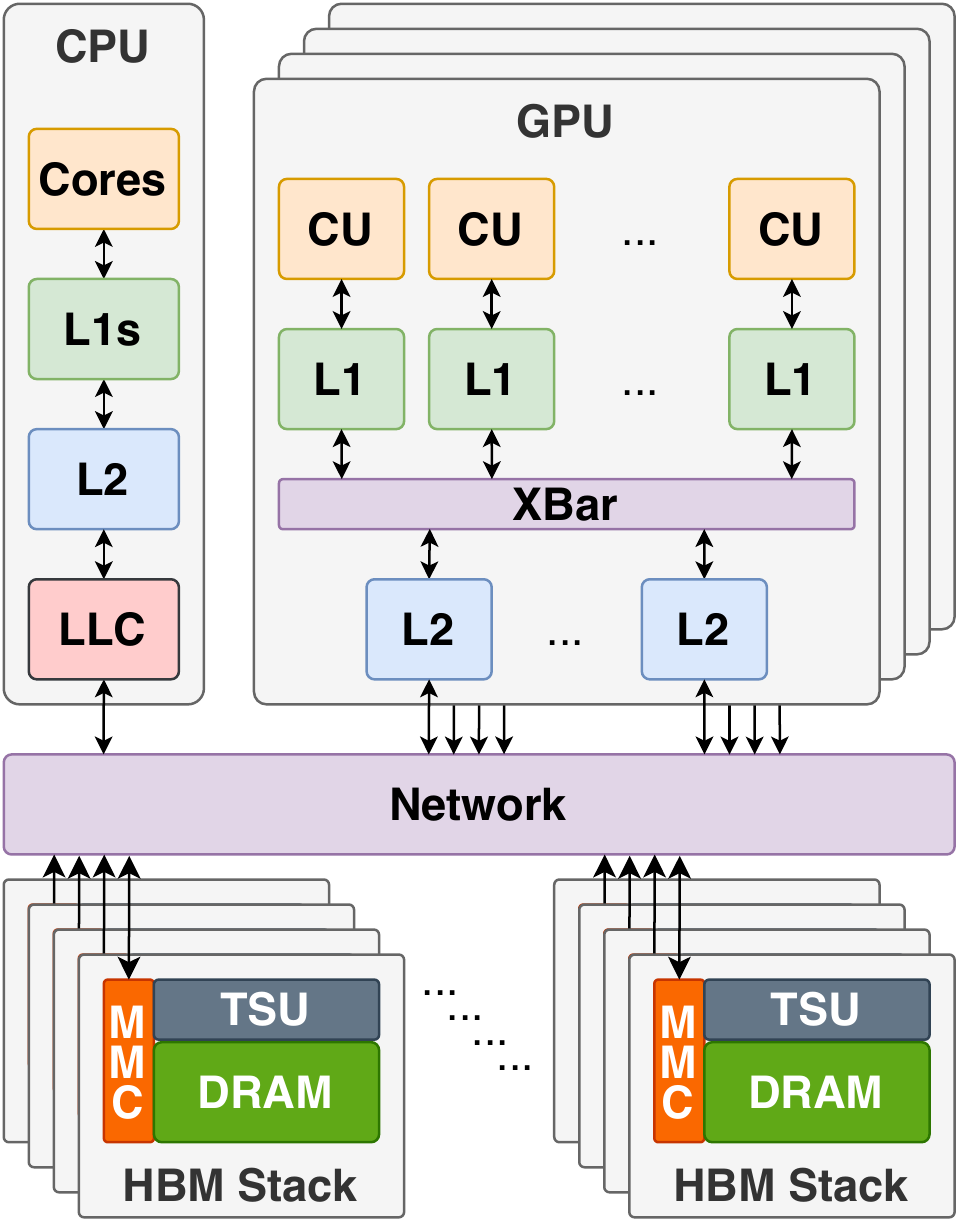}
    \caption{MGPU with shared main memory .}
    \label{fig:tsm1}
  \end{minipage}%
  \hspace{0.05\textwidth}
  \begin{minipage}{0.6\textwidth}
    \centering
    \includegraphics[width=\linewidth]{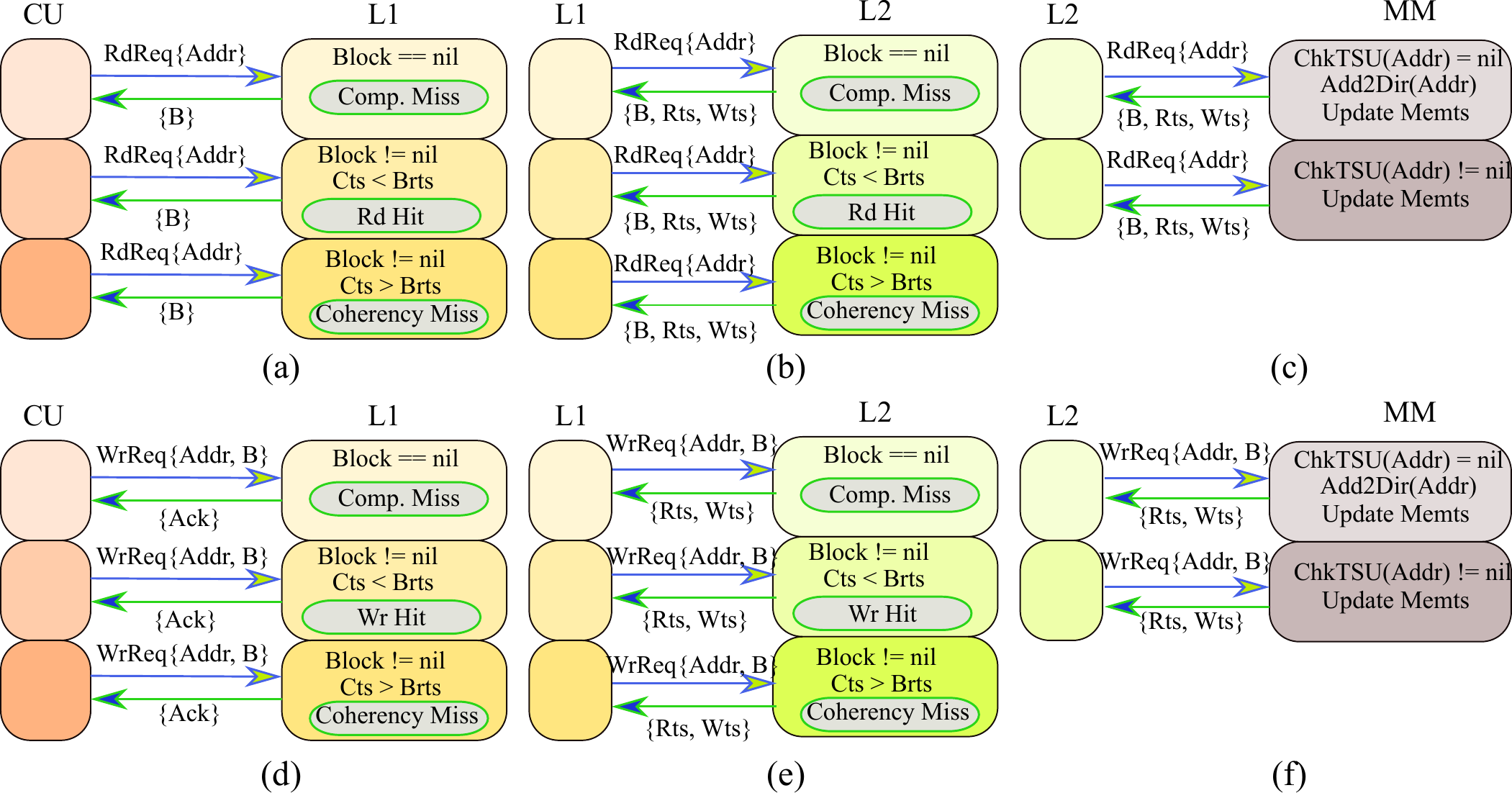}
    \caption{Request and response traffic between different components for each memory event. (a) Transactions between a CU and L1\$ for read operations, (b) Transactions between L1\$ and L2\$  for read operations, (c) Transactions between L2\$ and MM  for read operations, (d) Transactions between a CU and L1\$ for write operations, (e) Transactions between L1\$ and L2\$ for write operations, and (f) Transactions between L2\$ and MM for write operations}
    \label{fig:coh}
  \end{minipage}
  \vspace{-0.1in}
\end{figure*}
}
Figure~\ref{fig:tsm1} shows the logical organization of our target MGPU-SM system. In this configuration, each CU has a private L1\$ and each GPU has 8 distributed shared L2 banks. Each L2 bank has a corresponding cache controller (CC). We use High Bandwidth Memory (HBM) as the MM because HBM is power and area efficient, capable of providing high bandwidth required for GPUs~\cite{o2014highlights}. A network provides connectivity between the CC in the L2\$ and the MM controller (MMC) in the HBM. More details about the network are provided in Section~\ref{sec:method}. We assume a 4 GB memory per HBM stack in this example. Each L2 CC handles 2 GB (the size depends on the number of CCs and the HBM size) of the entire address space in the HBMs. Thus, all the GPUs are connected to all the HBM stacks, making the memory space physically shared across all GPUs. We use a TSU in each HBM to keep track of the timestamps for the blocks being accessed by the different L2\$s. We provide the detailed operation of the TSU later in this section.

\begin{figure*}[!htp]
  \centering 
  \includegraphics[width=\textwidth]{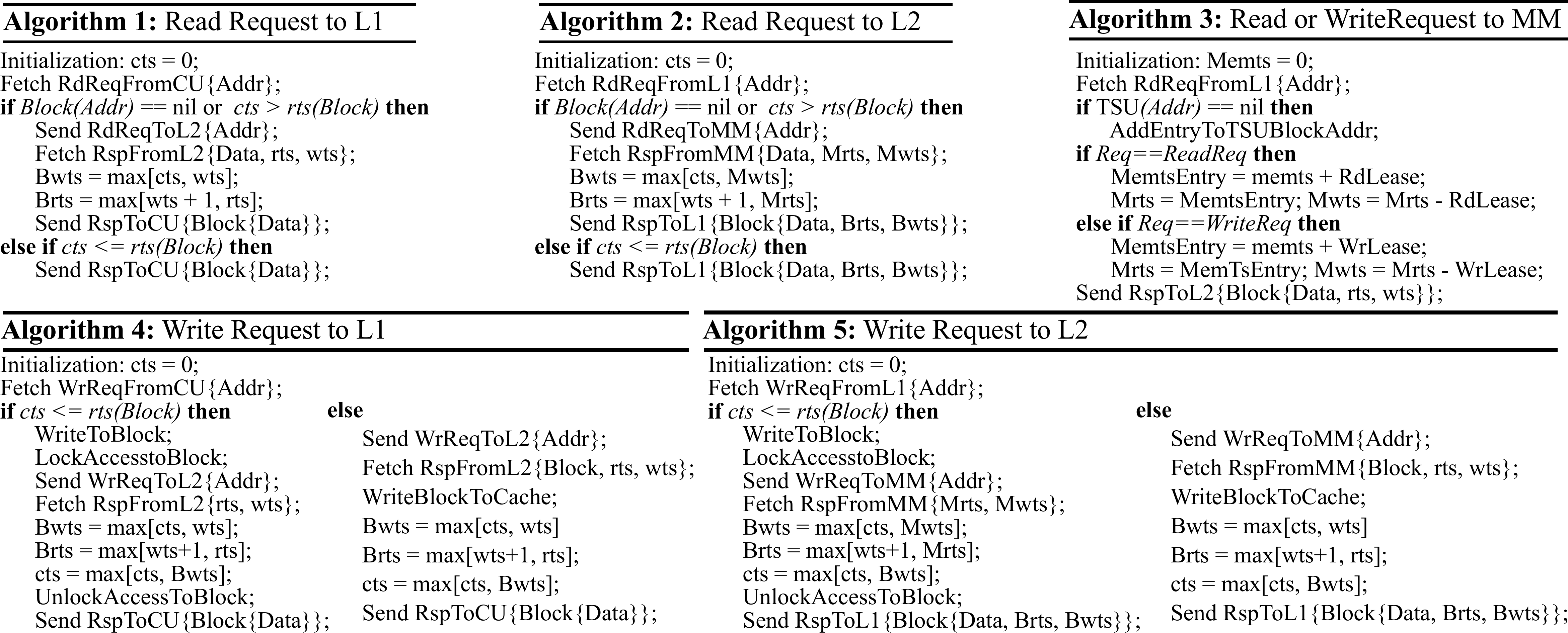}
  \vspace{-25pt}
\end{figure*}



\vspace{-3pt}
\subsection{HALCONE Protocol}
\label{subsec:halconeprotocol}
We define the \HALCONE protocol using a single-writer-multiple-reader (SWMR) invariant. Our HALCONE protocol is based on the G-TSC~\cite{tabbakh2018g} protocol designed for a single GPU. The terms used to explain the HALCONE protocol are defined in Table~\ref{tab:terms}. Unlike the G-TSC protocol, we do not have a \texttt{warpts} but assign a timestamp \texttt{cts} to each of the L1\$s and L2\$s. Each CU has a private L1\$, hence the \texttt{cts} for an L1\$  is equivalent to the \texttt{warpts} in the G-TSC protocol. Managing timestamps at the caches allows us to reduce timestamp traffic between the L1\$ and CU, as well as between the L1\$ and L2\$ by eliminating the need for sending \texttt{cts} with requests and responses to maintain coherence as compared to G-TSC protocol which sends \texttt{warpts} with every request. The memory operations are ordered based on the logical time, in particular, \texttt{cts}. If two requests have the same \texttt{cts} value, the cache uses physical time to order them. The key idea is that the block is only valid in the cache if the \texttt{cts} is within the valid \texttt{lease} period. Figure~\ref{fig:coh} shows the transactions between CUs, L1\$s, L2\$s, and MM for read and write operations. We explain these transactions with the help of Algorithms 1--5.

\vspace{-6pt}
\subsubsection{Read Operations}
\textbf{L1\$:} Figure~\ref{fig:coh}(a) shows the transactions between a CU and the L1\$ for read operations. As shown in Algorithm 1, a cache hit at L1\$ occurs only when there is an address (tag) match and the current timestamp, \texttt{cts}, is within the lease period of the cache block. If there is a tag hit, but the \texttt{cts} is not within the lease period, we fetch the cache block from L2\$ with new \texttt{rts} and \texttt{wts} values. For an L1\$ miss, we fetch the cache block with its \texttt{rts} and \texttt{wts} values from L2\$.

%

\textbf{L2\$:} Algorithm 2 shows how read requests are handled by the L2\$. The L2\$ hit or miss is similar to that of L1\$. Figure~\ref{fig:coh}(b) shows the transactions between the L1\$ and the L2\$ for read requests. If there is a cache hit and the lease is valid, the L2\$ sends the cache block, \texttt{rts}, and \texttt{wts} to the L1\$. If there is a cache miss in the L2\$, then the L2\$ sends a request to the MM. After fetching the cache block from MM, the L2\$ responds to the L1\$ with the cache block, \texttt{rts}, and \texttt{wts}. If there is a tag match, but \texttt{cts} is not within the lease period in L2\$, we re-fetch the data with new timestamps from the MM. This re-fetching of data ensures coherency in case another GPU modified the data in the MM. Note that G-TSC protocol only fetches renewed timestamps from L2\$ if data has not been modified by another CU. However, such re-fetching requires to send the \texttt{warpts} with each request (which we eliminated to reduce traffic) and adds more complexity as \HALCONE has a deeper memory hierarchy. 
\ignore{
one of two cases could have happened:\\
-- The lease expired, but the data has not been modified.\\
-- The lease expired, and data was modified by
another CU.\\
Irrespective of whether the data was modified,} 

%

\textbf{MM:} Figure~\ref{fig:coh}(c) shows the transactions to and from the MM for read requests from the L2\$. Algorithm 3 explains how a read request from the L2\$ is handled by the MMC. The MM tracks the timestamp of each block accessed by the L2\$s of all the GPUs using the TSU. The TSU stores the read address and the timestamp (\texttt{memts}) of the block, but not data itself. \texttt{memts} is used to keep track of the lease of a block sent to the L2\$s. If there is no entry for the requested address in the TSU (i.e., the block has never been requested by the L2\$s), it adds the address and then updates the \texttt{memts}\footnote{Read timestamp, \texttt{Mrts} and write timestamp, \texttt{Mwts} are design parameters for the \HALCONE protocol; depending on the implementation, these values can be staticcally or dynamically assigned.} of the block using the \texttt{Mrts } allocated for the read operation. If there is already an entry in the TSU for the requested address, the TSU extends the \texttt{memts} of the entry using the \texttt{Mrts} for the read operation.

\vspace{-6pt}
\subsubsection{Write Operations}
\textbf{L1\$:} Figure~\ref{fig:coh}(d) shows the transactions that take place for write requests to the L1\$. We adopt a write-through (WT)\footnote{We compared the run time of standard benchmarks using both L2\$ WT and L2\$ Write-back (WB) policies for an MGPU-SM system with no coherency. We observed that WT performs better for the MGPU-SM system. Hence, we implement our coherence protocol using L2\$ WT caches. For details on this, please refer to Section~\ref{sec:res}.} cache policy for both the L1\$s and L2\$s. Algorithm 4 illustrates how write requests to L1\$ are handled. Due to the WT policy, a write request to L1\$ triggers a write request from L1\$ to L2\$, irrespective of a cache hit or miss. If the \texttt{cts} is within the \texttt{lease}, it is a write hit. In case of a write hit, the data is written immediately to the cache block in the L1\$ and a write request is sent to the L2\$. Access to the block is locked until the L1\$ receives a write response, along with the new timestamps, from the L2\$. The access is locked by adding an entry to the miss-status-holding-register (MSHR). In the case of a write miss in the L1\$, the L1\$ sends the request to the L2\$. Once the L2\$ returns both the block and its timestamps to the L1\$, the L1\$ writes data to the appropriate location and updates its \texttt{cts}.

\textbf{L2\$:} Figure~\ref{fig:coh}(e) shows the transactions that take place for write requests to the L2\$s. Algorithm 5 demonstrates how a write request to the L2\$ is serviced. As we are using a WT policy for the L2\$, a write request to the L2\$ triggers a write request from the L2\$ to the MM, irrespective of whether the access is a cache hit or miss. Again, the cache hit and miss conditions are the same as in the case of the L1\$. If the access is a cache hit, the data is written to the block in the L2\$ and a write request is sent to the MM. The L2\$ updates the timestamp of the cache block using the response that it receives from the MM.  The access to the block is locked until the write response and the timestamps are received from the MM. If the L2\$ access results in a cache miss, L2\$ sends a write request to the MM. The write request includes the data and address. The MM sends a response with the block and updated timestamps. Next, the L2\$ issues a write to the block and updates its timestamps using the response from the MM.

\begin{figure*}[!htp]
  \centering
  \includegraphics[width=0.92\textwidth]{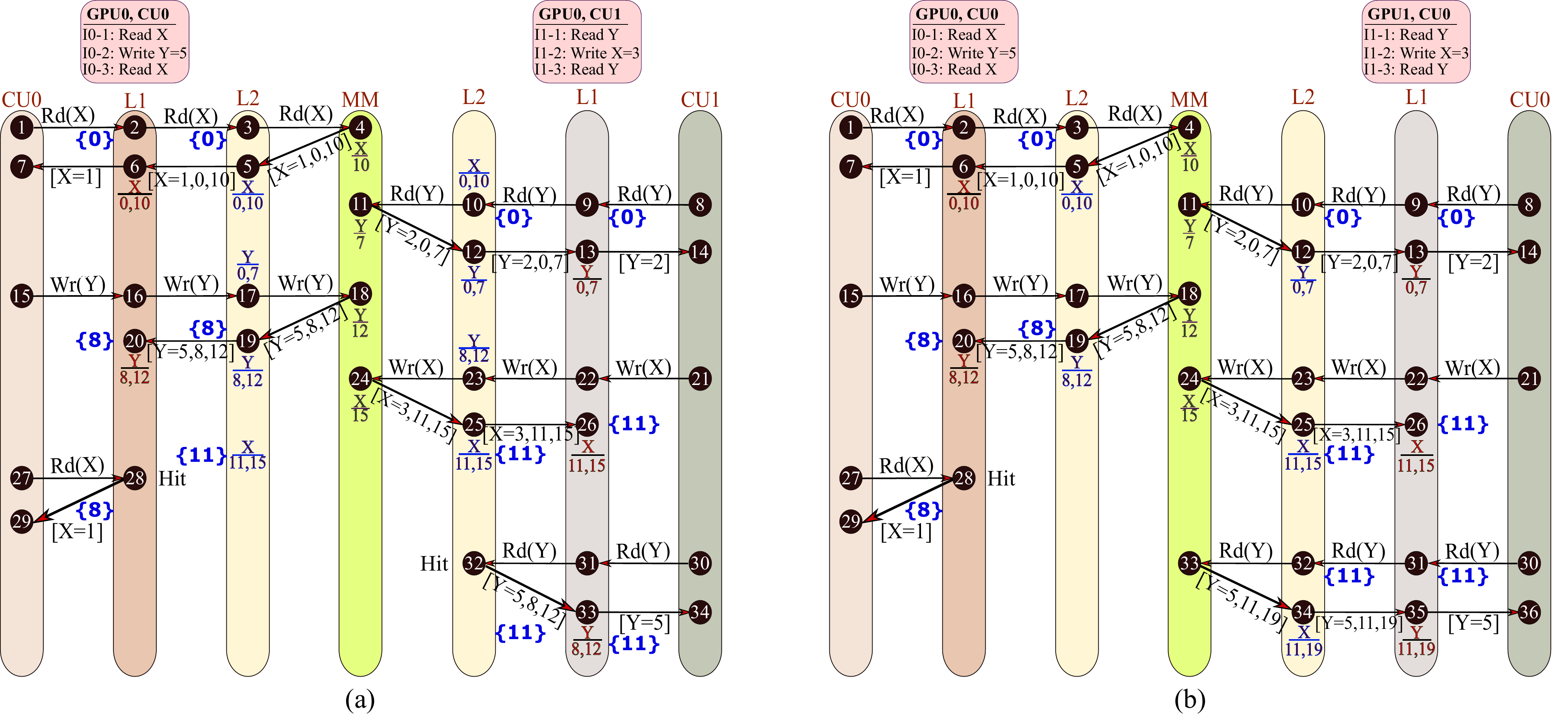} \vspace{-10pt}
  \caption{The timeline for (a) the intra- and (b) inter-GPU coherency. []
  represents response traffic in [Data, \texttt{rts}, \texttt{wts}] or [Data]
  format, \{\} represents the updated \texttt{cts} of a cache. In (a), the two L2\$ instances refer
  to the same physical L2\$.}
  \label{fig:time}
  \vspace{-13pt}
\end{figure*}
\textbf{MM:} Figure~\ref{fig:coh}(f) shows the transactions to and from the MM for a write request from the L2\$. Algorithm 3 explains how a write request from the L2\$ is serviced by the MMC. If there is no matching entry for the requested address in the TSU, then the TSU adds the address and updates the timestamp of the block using the \texttt{lease} for the write operation. If there is an entry present in the TSU for the requested address, the MM increases the \texttt{memts} of the entry using the \texttt{lease} for a write operation.

\vspace{-6pt}
\subsubsection{Intra-GPU Coherency} 
We use instructions identical to those described by Tabbakh et al.~\cite{tabbakh2018g} to explain both intra-GPU and inter-GPU coherency. Here, we first present how intra-GPU coherency is maintained using our \HALCONE protocol. Figure~\ref{fig:time}(a) shows the instructions and the sequence of steps for maintaining intra-GPU coherency. In this example, we have two compute units, CU0 and CU1. Both CU0 and CU1 belong to GPU0. Each CU has a private L1\$, but the L2\$ is shared between the two CUs. In Figure~\ref{fig:time}(a), we show two L2\$s for the sake of explanation, but both L2\$s are the same L2\$. CU0 executes 3 instructions, I0-1, I0-2, and I0-3, which read location [X], write to location [Y] and read location [X], respectively. Similarly, CU1 executes 3 instructions, I1-1, I1-2, and I1-3, which are: read location [Y], write location [X], and read location [Y], respectively. Both L1\$ and L2\$ have initial \texttt{cts} values of 0. \circled{1} to \circled{34} correspond to different memory events that occur during the execution of the three instructions. At \circled{1}, CU0 issues a read to location [X]. This request misses in the L1\$. At \circled{2}, so the L1\$ sends a read request to L2\$. As the request misses in L2\$ as well, the L2\$ sends a read request to the MM at \circled{3}. At \circled{4}, the MM sends the response to the L2\$ with \texttt{rts} and a \texttt{wts} values of 10 and 0, respectively (we choose these values of the timestamps for this example. Our protocol works correctly for any values of \texttt{rts} and \texttt{wts}). Based on the cache block and the timestamps received from MM, at \circled{5} L2\$ updates its \texttt{cts}, the block's \texttt{rts} and \texttt{wts}, and responds to L1\$ with the updated \texttt{rts} and \texttt{wts} values, along with the cache block. Similarly, the L1\$ updates its \texttt{cts}, and \texttt{rts} and \texttt{wts} values for the cache block at \circled{6}. The CU finally receives the data from L1\$ at \circled{7}.  Instruction I1-1 from CU1 issues a read from location [Y] and follows the same steps as I0-1. The CU1 receives the data through steps \circled{8} to \circled{14}. We assume a different lease (\texttt{wts}$=0$, \texttt{rts}$=7$) for location [Y] for this example.

CU0 requests to write to location [Y] at \circled{15}. The write request from a CU is served by the MM, regardless of whether it is a cache hit at L1\$ or L2\$. At \circled{16}, the L1\$ of CU0 sends a write request to L2\$. This results in a cache hit at L2\$ as the location [Y] was previously read by CU1 and \texttt{cts} $\le$ \texttt{rts}. At \circled{17}, the L2\$ sends a write request to the MM for location [Y]. The MM updates the value and timestamps for location [Y]. We assume a lease of 5 for write operations in this example. At \circled{18}, the MM sends the response with \texttt{rts}$=12$ and \texttt{wts}$=8$ for the block containing [Y] to the L2\$. Then the L2\$ updates the timestamps for [Y] and sets \texttt{cts}($=8$) at \circled{19} and sends the updated timestamps to the L1\$ of CU0. At \circled{20}, the L1\$ updates the timestamps for the block containing [Y] and the associated \texttt{cts}($=8$). Note that we do not show the actions to lock and unlock a block in the diagram for clarity. Every write request to a block in the cache must lock access to the block until receiving a response from the MM. At step \circled{21}, there is a write request (I1-2) from the CU1 at location [X]. This follows the same steps followed by I0-2. The response to the write request is executed in steps \circled{22} to \circled{26}.  Now, both L1\$ and L2\$ of CU1 have a \texttt{cts} value of 11 after completing the write request to location [X]. At \circled{27}, there is a read request for location [X] from CU0. At \circled{28}, the \texttt{cts} value is 8 and the block for location [X] has a \texttt{rts} value of 10. Hence, it is a cache hit in L1\$. Note that the advantage of using a logical timestamp is that it allows the scheduling of a memory operation in the future by assigning a larger \texttt{wts} value. Hence, the previous write on [X] by CU1 will be visible later to L1\$ of CU0 as it has a \texttt{cts} value lower than the assigned \texttt{wts} value to the block for the write request by CU1 at \circled{24}. Since the \texttt{cts} of the L1\$ of CUO is smaller than the \texttt{cts} value of the L1\$ of CU1 at this point, the read by CU0 of the L1\$ happens before the write by the L1\$ of CU1. The data is sent to CU0 by L1\$ at \circled{29}. At \circled{30}, CU1 sends a request to read location [Y]. This request creates a coherency miss in L1\$. This is because the \texttt{cts} is 11, but the block for location [Y] has a \texttt{rts} of 7. At \circled{31}, L1\$ sends a read request to L2\$. This request results in a cache hit at L2\$, since L2\$ has a \texttt{cts} value of 11 and the block for [Y] has \texttt{rts}$= 12$ and \texttt{wts}$=8$. The execution order of the instructions in this example is I0-1$\,\to\,$I1-1$\,\to\,$I0-2$\,\to\,$I0-3$\,\to\,$I1-2$\,\to\,$I1-3.

\vspace{-4pt}
\subsubsection{Inter-GPU Coherency}

In this example, we use the same instructions as in the previous example for intra-GPU coherency. CU0 of GPU0 executes instructions I0-1, I0-2, and I0-3. However, instructions I1-1, I1-2, and I1-3 are executed by the CU0 of GPU1 in this example. Thus, we have two different L2\$s, one connected to GPU0 and one connected to GPU1. Figure~\ref{fig:time}(b) shows the instructions and the sequence of execution for explaining inter-GPU coherency.  The read request from CU0 of GPU0 to read location [X] and read request from CU0 of GPU1 to read location [Y] follow the exact same steps (steps \circled{1} - \circled{14}) as in the case of intra-GPU coherency. The write request from CU0 of GPU0 at \circled{15} and the write request from CU0 of GPU1 at \circled{21} are also handled in the same manner as in the case of intra-GPU coherency. The only difference is that the data for the write of [X] and for the write of [Y] reside in different L2\$s. The read request (I0-3) by CU0 of GPU0 at \circled{27} still produces a cache hit in L1\$.  The read request issued by CU0 of GPU1 (I1-3) results in a different set of execution steps. This is because at \circled{32}, there is no longer an L2\$ hit, as the lease (\texttt{rts}$=7$, \texttt{wts}$=0$) expired for a \texttt{cts}$=11$. Hence the data for [Y] has to be fetched from the MM. The MM has the updated value written by CU0 of GPU0. This value is received by CU0 of GPU1, and thus it becomes coherent with CU0 of GPU0. The execution order for both the instructions in this example is again I0-1$\,\to\,$I1-1$\,\to\,$I0-2$\,\to\,$I0-3$\,\to\,$I1-2$\,\to\,$I1-3.

\vspace{-8pt}
\subsubsection{TSU Implementation}

\begin{figure}[t]
  \centering
  \includegraphics[width=0.7\columnwidth]{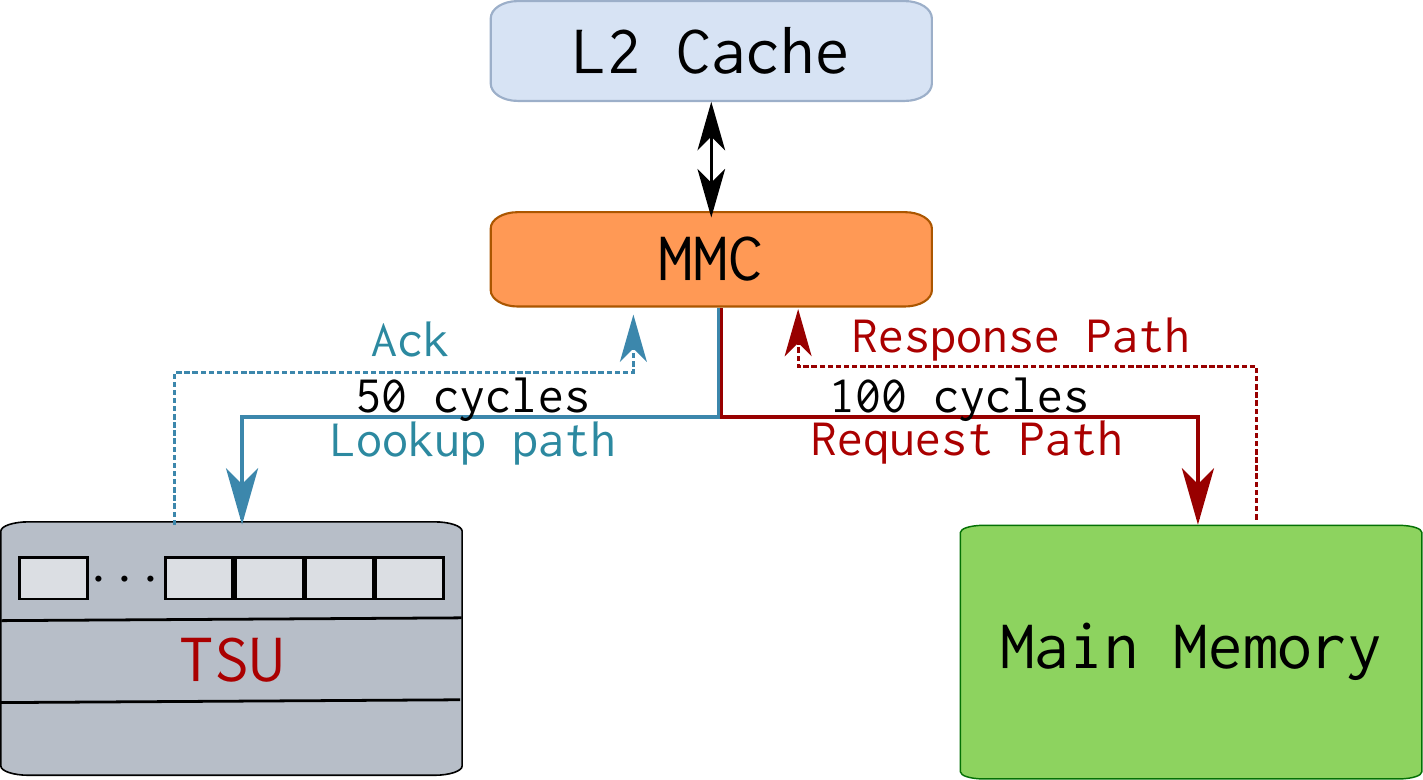}\vspace{-9pt}
  \caption{Time Stamp Unit (TSU). The TSU operates independently and in parallel with the memory access.}
  \label{fig:dir} \vspace{-17pt}
\end{figure}

The TSU is physically placed inside the logic layer of the HBM stack. We could have chosen to place the TSU in the DRAM layers, but this would increase memory access latency. We designed the TSU as an 8-way set associative cache. The TSU needs to store the \texttt{memts} for all of the blocks in all the L2\$s in the MGPU system. We use 16 bits for each \texttt{memts}. Since we have 8 distributed L2\$ modules in each GPU, each way of the TSU keeps track of the timestamps of the cache blocks in one of the L2\$ modules.  For example, for an MGPU with a 2MB L2\$ per GPU, we need 64KB of space for the timestamps in the TSU for each GPU. As TSU logic only searches for the presence of the timestamp of a block and generates or updates timestamps, the latency for accessing the TSU is identical to a L3\$  hit time of 40 cycles~\cite{levinthal2009performance}. We conservatively assume a 50 cycles access latency for TSU.

Figure~\mbox{~\ref{fig:dir}} shows the operation of the TSU. A request from the memory controller is sent to the TSU and the DRAM layer in parallel. The TSU responds with the timestamp for the cache block, and in parallel with the DRAM layer, responds with the cache block. Thus, the TSU never impacts the critical path of the DRAM access, and so does not add any performance overhead. The eviction of TSU entries is related to the eviction of L2\$ entries. When there is an eviction from the L2\$ of a GPU, the TSU also evicts the timestamp for that cache block if it is not shared with other GPUs. The TSU logic determines the block sharers using the \texttt{memts} value (if the value of \texttt{memts} is within one lease period, it is assumed to be shared). In case the TSU is full, the TSU evicts the cache block with lowest \texttt{memts} value.  

\vspace{-5pt}
\subsubsection{Timestamp Design}
We use 16-bit fields for each one of the timestamps, \texttt{rts} and \texttt{wts}. Assuming 64B cache block size, 4B for ACK, 4B for metadata and 8B address, \HALCONE increases the network traffic by 5\% and 5.26\% for read and write transactions, respectively. If the timestamp value overflows, instead of flushing the cache, we simply re-initialize the timestamps to 0. This re-initialization results in a cache miss for one of the cache blocks. However, given we are using a write-through policy for writes in both L1\$ and L2\$, there is no chance of losing data belonging to the cache block experiencing the overflow. 
We just need to do an extra MM access. We need 1KB of storage per L1\$ of size 256 KB and 128 KB of storage per L2\$ of size 2 MB for holding the read and write timestamps. For each cache timestamp (\texttt{cts}), we use 64 bits. For an example GPU with 32 CUs, the GPU requires a total of 40 \texttt{cts} entries (32 for the 32 private L1\$s belonging to each CU and 8 for the L2\$). Hence, we need a total of 320 bytes to represent all the \texttt{cts} values for the entire GPU.

\ignore{
\subsubsection{G-TSC vs. HALCONE}
\Com{Although our work builds on top of the G-TSC protocol, our HALCONE protocol has several novelties. The comparison between G-TSC and HALCONE is as follows:
}
\newline
a) \Com{G-TSC is designed for intra-GPU coherency and cannot be readily applied
for multi-GPU coherency. Maintaining coherency across multiple GPUs is more
challenging as the L1\$s of a GPU can only interact with its own L2\$.
We need to maintain coherency across multiple L2\$s across different GPUs. A
straightforward extension of G-TSC would require inclusion of timestamps for
each of the blocks in the MM, leading to signficant area overhead.
Hence, we introduce TSU support near the MM to reduce performance
and area overhead.} \newline b) \Com{Our use of TSU is unique in the sense
that it has no additional performance overhead because of its strategic
positioning and operating principles.} \newline c) \Com{We eliminate the
{\texttt{wts}} and use {\texttt{cts}} to reduce request traffic by 41.7\% in
comparison to G-TSC.} \newline d) \Com{While G-TSC simply provides the same
lease for both reads and writes, we provide different lease values for reads and
writes. By doing so, we favor temporal locality behavior. This is again a unique
feature compared to G-TSC.}}

\vspace{-10pt}
\section{Evaluation Methodology}
\label{sec:method}

In this section, we describe the MGPU system configurations, the simulator, standard application benchmarks, and the custom synthetic benchmarks (for stress testing \HALCONE) that are used to evaluate different MGPU configurations.

\vspace{-5pt}
\subsection{MGPU System Configurations}
Table~\ref{tab:configuration} shows the architecture of each GPU in our MGPU system. To complete a comprehensive evaluation, we evaluate five different MGPU configurations\footnote{To name the MGPU systems we use the following notation: \texttt{C} = cache-coherence support, \texttt{NC} = No coherency support, \texttt{WT} = L2\$ with write-through policy, and \texttt{WB} = L2\$ with write-back policy.}: 
\vspace{-0.075in}
\begin{enumerate}
  \item MGPU system with RDMA (\texttt{RDMA-WB-NC}). 
  \vspace{-0.075in}
  \item MGPU system with RDMA and HMG coherency\footnote{HMG is the most recently proposed solution for efficient HW cache-coherent support in MGPU systems (HMG is succinctly described in Sections~\ref{sec:introduction} and~\ref{sec:related}, and with more details in Section~\ref{sec:method}).} (\texttt{RDMA-WB-C-HMG}). 
  \vspace{-0.075in}
  \item MGPU-SM system, L2\$ with WB and no coherency (\texttt{SM-WB-NC}).  
  \vspace{-0.075in}
  \item MGPU-SM system, L2\$ with WT and no coherency (\texttt{SM-WT-NC}).
  \vspace{-0.075in}
  \item MGPU-SM system, L2\$ with WT and \HALCONE (\texttt{SM-WT-C-\HALCONE}).  
\end{enumerate} 
\vspace{-0.075in}

\begin{table}[tp]
  \centering 
  \caption{GPU Architecture.}
  \resizebox{0.49\textwidth}{!}{
  \begin{tabular}{@{}lll|llr@{}}
  \toprule

  \textbf{Component} & \textbf{Configuration} & \textbf{Count} &
  \textbf{Component} & \textbf{Configuration} & \textbf{Count}\\

  \textbf{per GPU} & &  & \textbf{per GPU}&\\ \midrule CU & 1.0 GHz & 32 & L1 Vector \$ & 16KB 4-way & 32\\

  L1 Scalar \$ & 16KB 4-way & 8 & L1I\$ & 32KB 4-way & 8\\

  L2\$ & 256KB 16-way & 8 & DRAM & 512MB HBM & 8 \\

  L1 TLB & 1 set, 32-way & 48 & L2 TLB & 32 sets, 16-way & 1\\

  \bottomrule
  \end{tabular}} \vspace{-15pt}
  \label{tab:configuration}
\end{table}

Figure~\ref{fig:nvlink} shows the {\texttt{RDMA}} configuration for a typical MGPU system. In this configuration, each GPU's L1\$ is connected to a switch (SW) that connects to another GPU's L2\$ for RDMA. Each switch forwards 16 bits per transfer. Switches run at 16 GTransfers/s. Thus, each switch can support a throughput of 32 GB/s (unidirectional link bandwidth between L2 and MM), which is the peak unidirectional bandwidth for PCIe 4.0~\cite{gonzales2015pci}. In the case of HMG, to maintain coherency, each GPU's L2\$ is connected to the switch (SW) and the protocol uses RDMA via L2\$.  For RDMA connections between L2\$ and MM, we use PCIe 4.0 links. For both baseline and HMG configurations, the MGPUsim simulator faithfully models the PCIe interconnects. For our MGPU-SM system, we group together the switches to form a switch complex. Both the L2\$s and the MM are connected to the switch complex. The overall L2-to-MM bidirectional bandwidth is 256 GB/s, though each HBM supports an effective communication bandwidth of 341 GB/s~\cite{cho20181}. Hence, in our MGPU-SM evaluation, the total L2-to-MM bandwidth is limited to 1 TB/s. We carefully model the queuing latency on the L2-to-MM network, as well as a fixed 100-cycle latency at the memory controllers (the number is calibrated using a real GPU with HBM memory). For our evaluation, we allocate memory by interleaving 4 KB pages across all the memory modules in the MGPU system.

Our evaluation of the \texttt{RDMA-WB-NC} and \texttt{SM-WB-NC} configurations is aimed at exposing the need for our proposed MGPU-SM cache coherent systems (more details in Section~\ref{sec:res}). The \texttt{SM-WB-NC} and \texttt{SM-WT-NC} configurations are used to compare L2\$ write-back (WB) policy with L2\$ write-through (WT) policy in a MGPU-SM system. This comparison helps us learn which write policy is more suitable for L2\$ in a MGPU-SM system. The \texttt{SM-WT-NC} and \texttt{SM-WT-C-\HALCONE}configurations are then compared to determine the overhead of coherency (we use a WT policy as it provides better performance than WB, as reported in our experiments in Section~\ref{sec:res}). 
The comparison between configuration \texttt{RDMA-WB-C-HMG} and \texttt{SM-WT-C-\HALCONE} demonstrates the improvement achieved by our proposed solution over the most optimized hardware coherence support for MGPU systems (HMG protocol). Except for HMG which leverages scope based memory consistency model, we adopt existing weak memory consistency model for our evaluation. Nonetheless, our \HALCONE protocol can work as a building block for more strict memory consistency models. 

\ignore{For the \texttt{RDMA-WB-NC} and \texttt{RDMA-WB-C-HMG} configurations, we provide RDMA support for peer-to-peer access among the GPUs. For the shared memory configurations (\texttt{SM-}), we do not need RDMA support because all the GPUs must have access to all the memory modules. We design a network that provides connectivity through the switch-complex between every GPU-memory module pair.} 

\vspace{-5pt}
\subsection{Simulation Platform} 
We use the MGPUSim~\cite{sun2019mgpusim} simulator to  model MGPU systems. MGPUSim has been validated against real AMD MGPU systems. 
We modified the simulator and its memory hierarchy to support \HALCONE. After implementing the \HALCONE protocol, we verify the implementation using unit, integration, and acceptance tests provided with the simulator. We also modified the simulator to support the HMG protocol by implementing a hash function that assigns a home node for a given address, directory support for tracking sharers and invalidation support for sending messages to the sharers as needed.

\vspace{-5pt}
\subsection{Benchmarks} 
We use standard application GPU benchmarks as well as synthetic benchmarks to evaluate our \HALCONE protocol in an MGPU-SM system.

\vspace{-5pt}
\subsubsection{Standard Benchmarks}
\label{sec:regb} 
We use a mix of memory-bound and compute-bound benchmarks, 11 in total (see Table~\ref{tab:ben}), from the Hetero-Mark~\cite{heteromark}, PolyBench~\cite{pouchet2012polybench}, SHOC~\cite{shoc}, and DNNMark~\cite{dong2017dnnmark} benchmark suites to examine the impact of our \HALCONE protocol on the MGPU-SM system. In addition, these workloads have large memory footprints and represent a variety of data sharing patterns across different GPUs. More details about the benchmarks can be found in~\cite{sun2018mgsim+,sun2019mgpusim}.

\begin{table}[t] 
\centering
\caption{Standard benchmarks used in this work.
Memory represents the footprint in the GPU memory.}
\resizebox{\columnwidth}{!}
{\begin{tabular}{@{}lllr@{}} 
\toprule
\textbf{Benchmark (abbr.)} & \textbf{Suite} & \textbf{Type} & \textbf{Memory} \\
\midrule Advanced Encryption    & \multirow{2}{*}{Hetero-Mark} & \multirow{2}{*}{Compute} & \multirow{2}{*}{71 MB}\\ 
Standard (\textbf{aes}) & & &\\ 
Matrix Transpose and    & \multirow{2}{*}{PolyBench} & \multirow{2}{*}{Memory} & \multirow{2}{*}{64 MB}\\ Vector Multiplication (\textbf{atax}) & & & \\
Breadth First Search (\textbf{bfs}) & SHOC & Memory  & 574 MB\\ 
BiCGStab Linear Solver (\textbf{bicg})   & PolyBench & Compute & 64 MB\\
Bitonic Sort (\textbf{bs}) & AMDAPPSDK & Memory  &  67 MB \\
Finite Impulse Response (\textbf{fir}) & Hetero-Mark & Memory & 67 MB\\
Floyd Warshall (\textbf{fws}) & AMDAPPSDK & Memory &  32 MB\\
%
%
Matrix Multiplication (\textbf{mm}) & AMDAPPSDK & Memory &  192 MB\\
%
%
Maxpooling (\textbf{mp})   & DNNMark & Compute & 64 MB \\
%
%
Rectified Linear Unit (\textbf{rl})   & DNNMark & Memory & 67 MB \\
Simple Convolution (\textbf{conv}) & AMDAPPSDK & Memory & 145 MB\\
%
\bottomrule
\vspace{-30pt}
\end{tabular}
\label{tab:ben} }
\label{benchmarks}
\end{table}

\vspace{-5pt}
\subsubsection{Synthetic Benchmarks} 
The publicly available benchmark suites mentioned in Section~\ref{sec:regb} have been developed considering the lack of hardware-level coherency support in GPUs. Hence, these benchmarks cannot necessarily harness the potential benefit of the hardware-support for coherency in our MGPU-SM system. To stress test our \HALCONE protocol, we develop a synthetic benchmark suite called \textbf{Xtreme}. There are three benchmarks in the Xtreme suite\footnote{All the benchmarks in the \textbf{Xtreme} suite perform repeated writes to and reads from the same location. This extreme behavior is typically uncommon in regular GPU benchmarks, and so the name Xtreme.}. All the benchmarks in the Xtreme suite perform a basic vector operation: $\mathrm{C = A + B}$, where $\mathrm{A}$ and $\mathrm{B}$ are floating point vectors. We describe the basic operation of the \textbf{Xtreme} benchmarks with a simple example. For each example, we assume the following:
\vspace{-0.05in}
\begin{enumerate} 
  \item There are two GPUs: GPU\textit{$_X$} and GPU\textit{$_Y$}.
  \vspace{-0.075in}
  \item Both GPU\textit{$_X$} and GPU\textit{$_Y$} are equipped with two CUs each: CU\textit{$_{X0}$}, CU\textit{$_{X1}$}, and CU\textit{$_{Y0}$} and CU\textit{$_{Y1}$}, respectively. 
  \vspace{-0.075in}
  \item There are three vectors $\mathrm{A}$, $\mathrm{B}$ and $\mathrm{C}$ that are used to compute $\mathrm{C = A + B}$ using both GPU\textit{$_X$} and GPU\textit{$_Y$}.
  \vspace{-0.075in}
  \item Each of the three vectors, $\mathrm{A}$, $\mathrm{B}$ and $\mathrm{C}$, are split into 4 slices: $\mathrm{A_0}$, $\mathrm{A_1}$, $\mathrm{A_2}$ and $\mathrm{A_3}$; $\mathrm{B_0}$, $\mathrm{B_1}$, $\mathrm{B_2}$ and $\mathrm{B_3}$; and $\mathrm{C_0}$, $\mathrm{C_1}$, $\mathrm{C_2}$, and $\mathrm{C_3}$.  
  \vspace{-0.075in}
  \item At the beginning of the program,
  CU\textit{$_{X0}$} reads $\mathrm{A_0}$, $\mathrm{B_0}$, and $\mathrm{C_0}$; CU\textit{$_{X1}$} reads $\mathrm{A_1}$, $\mathrm{B_1}$, and $\mathrm{C_1}$; CU\textit{$_{Y0}$} reads $\mathrm{A_2}$, $\mathrm{B_2}$, and $\mathrm{C_2}$; CU\textit{$_{Y1}$} reads $\mathrm{A_3}$, $\mathrm{B_3}$, and $\mathrm{C_3}$.
\end{enumerate} 
\vspace{-0.05in}
The three \textbf{Xtreme} benchmarks work as follows: \newline
\underline{\textbf{Xtreme1:}} \newline 
\circld{1} CU\textit{$_{X0}$} performs $\mathrm{C_0=A_0+B_0}$; Similarly, CU\textit{$_{X1}$} operates on $\mathrm{A_1}$, $\mathrm{B_1}$ and $\mathrm{C_1}$; CU\textit{$_{Y0}$} operates on $\mathrm{A_2}$, $\mathrm{B_2}$ and $\mathrm{C_2}$; and CU\textit{$_{Y1}$} operates on $\mathrm{A_3}$, $\mathrm{B_3}$ and $\mathrm{C_3}$.\\
\circld{2} Repeat step \circld{1} 10 times.\\ 
\circld{3} CU\textit{$_{X0}$} performs $\mathrm{A_0=C_0+B_0}$; Similarly, CU\textit{$_{X1}$} operates on $\mathrm{A_1}$, $\mathrm{B_1}$ and $\mathrm{C_1}$; CU\textit{$_{Y0}$} operates on $\mathrm{A_2}$, $\mathrm{B_2}$ and $\mathrm{C_2}$; and CU\textit{$_{Y1}$} operates on $\mathrm{A_3}$, $\mathrm{B_3}$ and $\mathrm{C_3}$.\\
\circld{4} Repeat step \circld{3} 10 times.\\ 
With \textbf{Xtreme1}, we evaluate the impact of consecutive writes to the same location by a CU. There is no data sharing between the CUs or the GPUs. When there is a write to any location, the corresponding \texttt{cts} of the L1\$ and L2\$ step ahead and generate read misses. Steps \circld{2} and \circld{4} force coherency misses in the caches. 
\newline 
\underline{\textbf{Xtreme2:}} \newline 
\circld{1} CU\textit{$_{X0}$} performs $\mathrm{C_0=A_0+B_0}$; Similarly, CU\textit{$_{X1}$} operates on $\mathrm{A_1}$, $\mathrm{B_1}$ and $\mathrm{C_1}$; CU\textit{$_{Y0}$} operates on $\mathrm{A_2}$, $\mathrm{B_2}$ and $\mathrm{C_2}$; and CU\textit{$_{Y1}$} operates on $\mathrm{A_3}$, $\mathrm{B_3}$ and $\mathrm{C_3}$.\\ 
\circld{2} CU\textit{$_{X0}$} performs $\mathrm{A_1=C_1+B_1}$;\\ 
\circld{3} Repeat step \circld{2} 10 times.\\ 
\circld{4} Repeat step \circld{1}\\ 
With \textbf{Xtreme2}, we stress test \HALCONE for intra-GPU coherency. There is a SWMR invariant dependency between CU\textit{$_{X0}$} and CU\textit{$_{X1}$} at \circld{2}, CU\textit{$_{X0}$} writes to a location that was previously read by CU\textit{$_{X1}$}. Step \circld{3} forces coherency misses. \newline 
\underline{\textbf{Xtreme3:}} \newline 
\circld{1} CU\textit{$_{X0}$} performs $\mathrm{C_0=A_0+B_0}$; Similarly, CU\textit{$_{X1}$} operates on $\mathrm{A_1}$, $\mathrm{B_1}$ and $\mathrm{C_1}$; CU\textit{$_{Y0}$} operates on $\mathrm{A_2}$, $\mathrm{B_2}$ and $\mathrm{C_2}$; and CU\textit{$_{Y1}$} operates on
$\mathrm{A_3}$, $\mathrm{B_3}$ and $\mathrm{C_3}$.\\ 
\circld{2} CU\textit{$_{X0}$} performs $\mathrm{A_3=C_3+B_3}$;\\ 
\circld{3} Repeat step \circld{2} 10 times.\\ 
\circld{4} Repeat step \circld{1}\\ 
With \textbf{Xtreme3}, we stress test \HALCONE for inter-GPU coherency. The difference between \textbf{Xtreme2} and \textbf{Xtreme3} is that at \circld{2} CU\textit{$_{X0}$} writes to a location that was previously read by CU\textit{$_{X1}$} and CU\textit{$_{Y1}$}, respectively.

In our evaluation, we vary vector sizes from 192 KB to 96 MB for $\mathrm{A}$, $\mathrm{B}$ and $\mathrm{C}$ so that we can examine the impact of capacity misses at different levels of the memory hierarchy.

\vspace{-8pt}
\section{Evaluation}
\vspace{-2pt}
\label{sec:res}

\begin{figure*}[!htp]
	\centering \vspace{-2pt}
    \subfigure{\label{fig:reg-speedup}\includegraphics[width=0.325\textwidth]{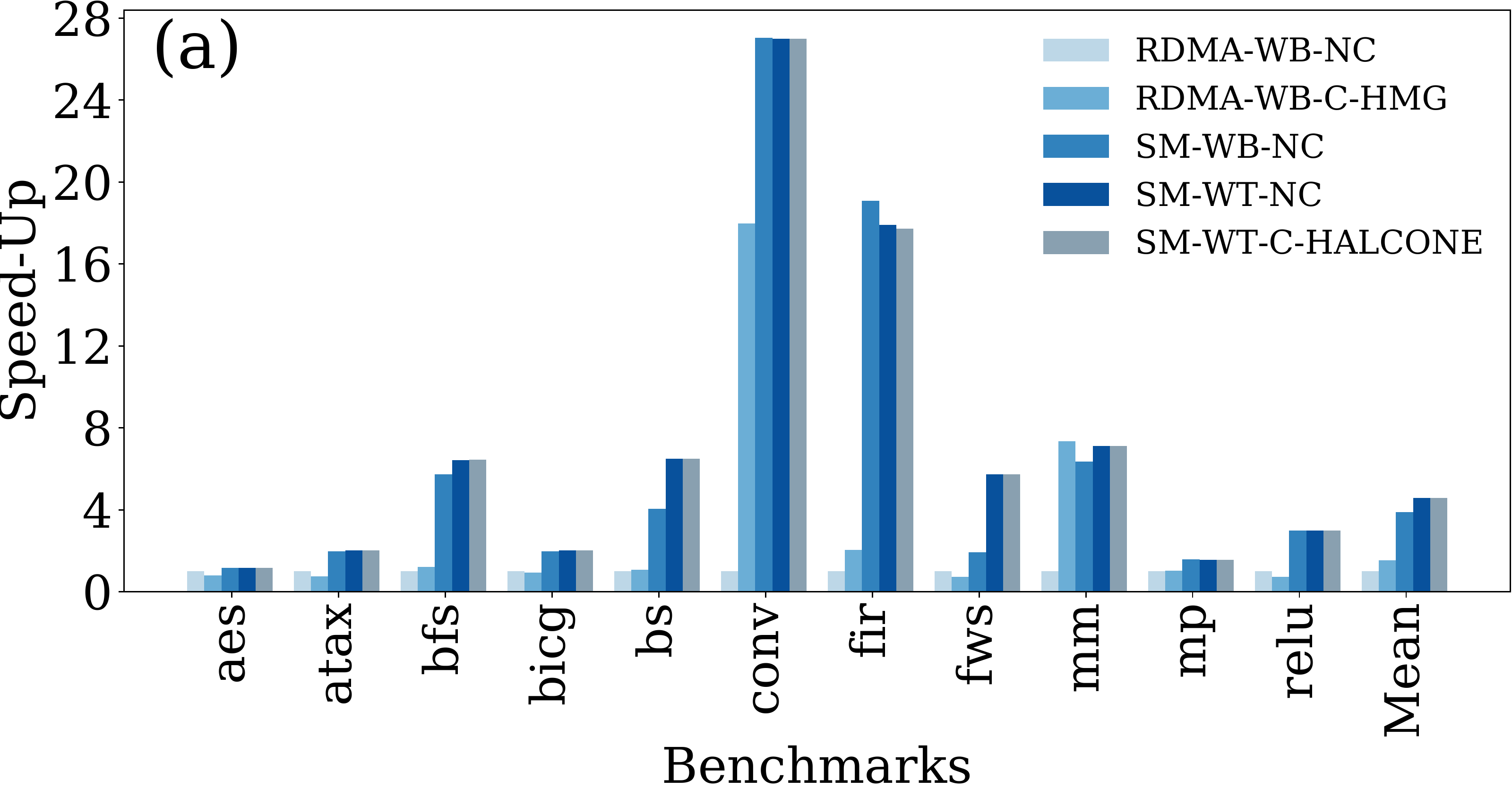}}
	\subfigure{\label{fig:L2-MM}\includegraphics[width=0.325\textwidth]{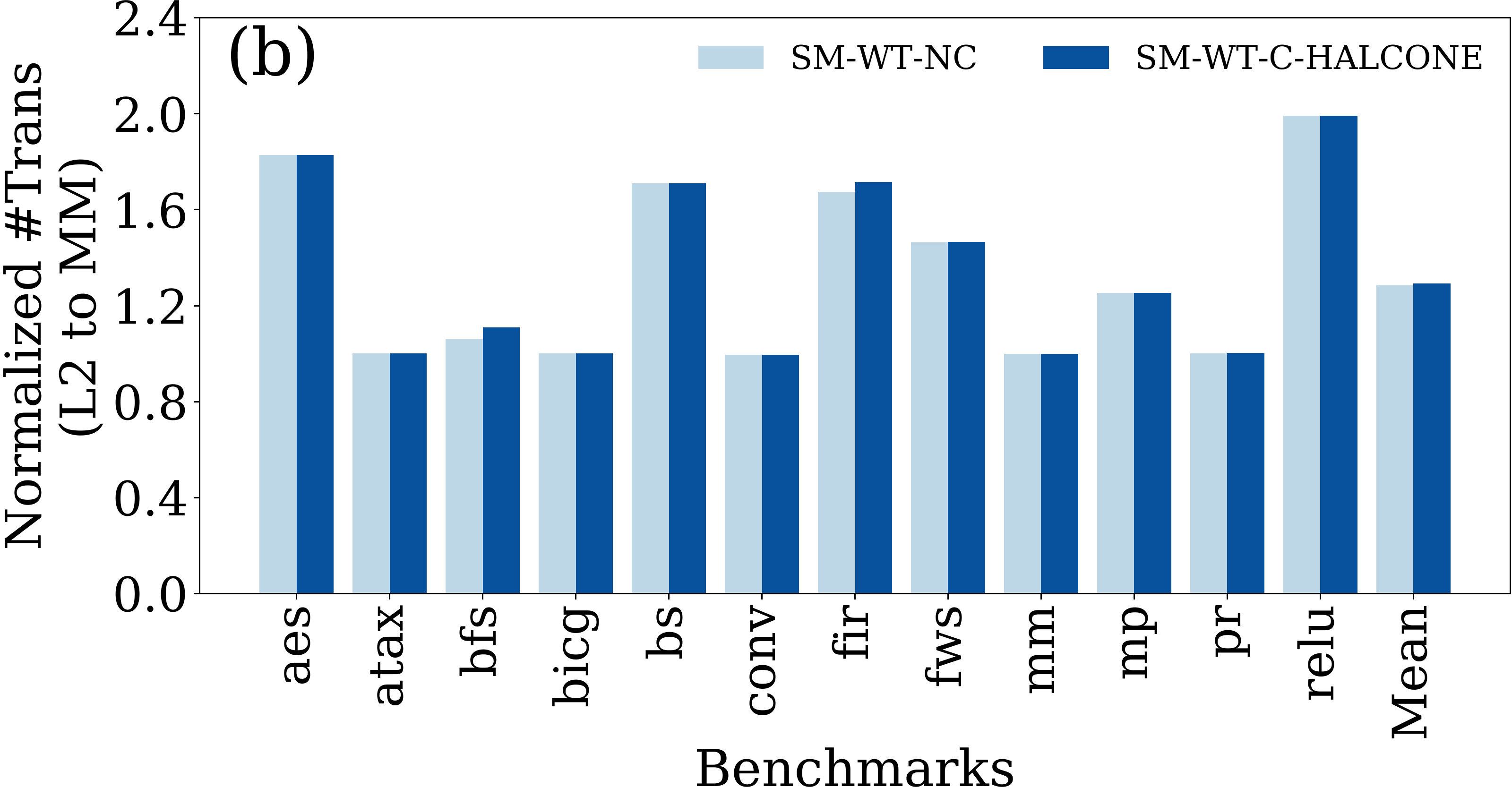}}
	\subfigure{\label{fig:L1-L2}\includegraphics[width=0.325\textwidth]{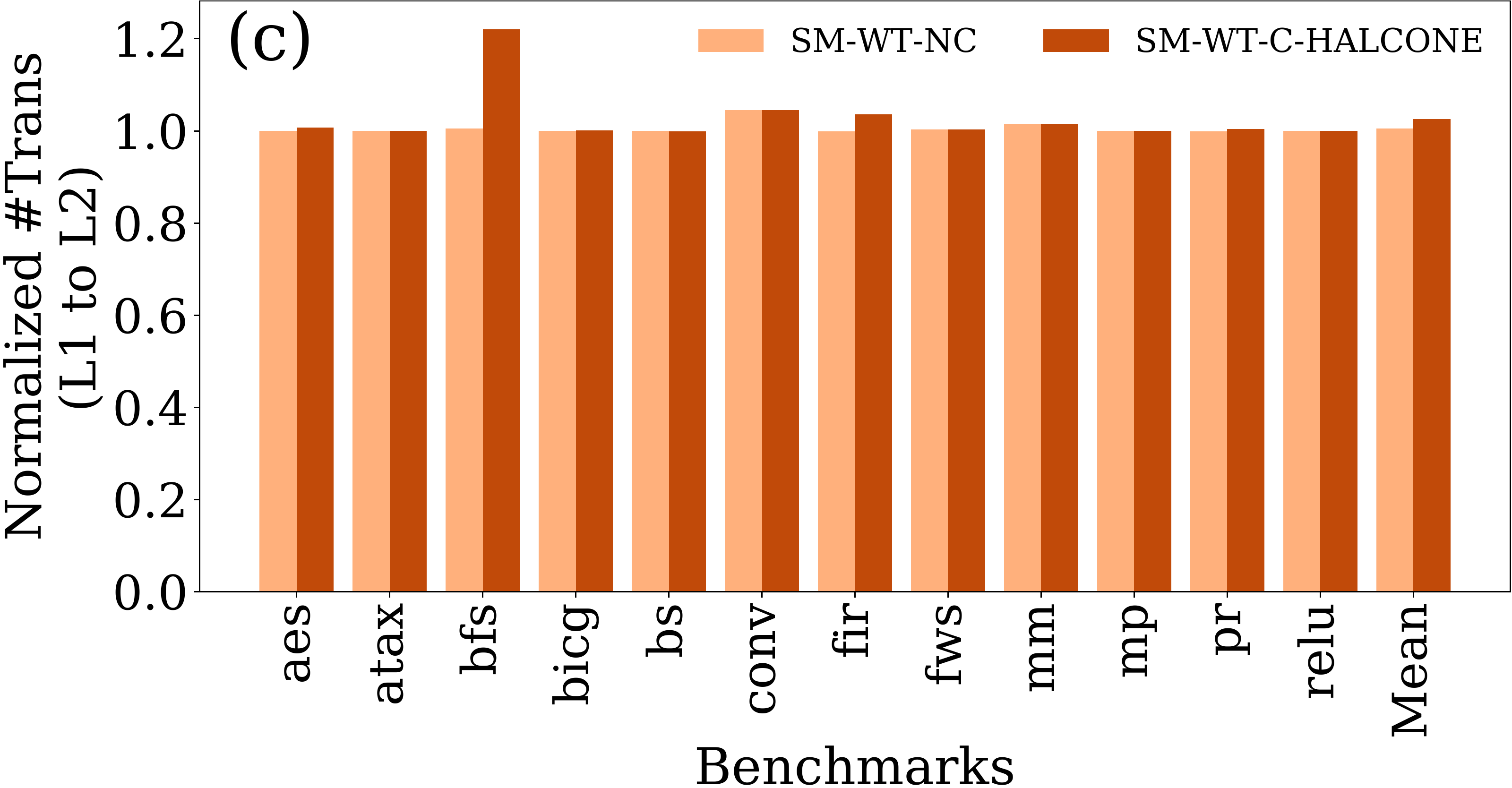}} \vspace{-8pt}
	\caption{(a) Speed-up for different MGPU systems, normalized versus \texttt{RDMA-WB-NC}. (b) Number of L2\$ to MM transactions for \texttt{SM-WT-NC} and \texttt{SM-WT-C-\HALCONE} normalized  versus the number of transactions for the \texttt{SM-WB-NC} configuration in an MGPU system. (c) Number of L1\$ to L2\$ transactions for \texttt{SM-WT-NC} and \texttt{SM-WT-C-\HALCONE}normalized  versus the number of transactions for the \texttt{SM-WB-NC} configuration in an MGPU system. Mean refers to geometric mean.}
  \label{fig:tsm_results}
	\vspace{-0.22in}
\end{figure*}

In this section, we present our evaluation of the \HALCONE protocol for both existing standard benchmarks and synthetic benchmarks. The standard benchmarks have been developed in accordance with the current MGPU programming model that assumes no hardware support for coherency and places the burden of maintaining coherency on the programmer. Hence, we evaluate these traditional GPU benchmarks to ensure our \HALCONE protocol does not introduce extra overhead for these legacy cases, where coherency has been maintained by the GPU programmer. Next, we evaluate the \textbf{Xtreme} benchmarks, a set of three synthetic benchmarks that leverage hardware support for coherency to ensure the correctness of their computations.

\vspace{-6pt}
\subsection{Standard Benchmarks} \label{sec:regular}
We compare 5 different MGPU configurations: \texttt{RDMA-WB-NC} is our baseline, \texttt{RDMA-WB-C-HMG}, \texttt{SM-WB-NC}, \texttt{SM-WT-NC} and \texttt{SM-WT-C-\HALCONE} assuming a 4-GPU system. We use a \texttt{WrLease} of 5 and
a \texttt{RdLease} of 10 for this evaluation. Please refer to Section~\ref{sec:ts} for details on why we choose these lease values.


\ignore{and L2\$ latency\footnote{L2\$ latency is a measure of average latency for all cache accesses. Increase in cache latency is an indicator of increased cache misses. It is very important to note that L2\$ latency only accounts for latency for serving a request using a GPU's own memory, it does not account for the RDMA latency}}

Figure~\ref{fig:reg-speedup} shows the speed-up for different MGPU configurations, as compared to \texttt{RDMA-WB-NC}. Our evaluation shows that the \texttt{RDMA-WB-C-HMG}, \texttt{SM-WB-NC}, \texttt{SM-WT-NC}, and \texttt{SM-WT-C-\HALCONE} configurations achieve, on average, a 1.5$\times$, 3.9$\times$, 4.6$\times$, and 4.6$\times$ speed-up, respectively, versus \texttt{RDMA-WB-NC}. There are two reasons why all 3 shared memory configurations are faster than using \texttt{RDMA-WB-NC} alone. First, \texttt{RDMA-WB-NC} requires data copy operations between the CPU and GPUs. Shared memory eliminates this traffic since the CPU and GPUs share the same memory. Second, during kernel execution, all of the GPUs are required to use RDMA to access data residing on other GPUs' memory for the baseline. The shared main memory allows sharing of data across GPUs with no RDMA overhead.

For the compute--bound benchmarks (i.e., \textbf{aes}, \textbf{atax},
\textbf{bicg}, and \textbf{mp}), all the MGPU-SM configurations achieve lower (1.2$\times$ to 2.0$\times$) speed-up as compared to the speed-up achieved for the memory-bound benchmarks (3$\times$ to 27$\times$). This is due to the memory-bound benchmarks' higher reliance on the high overhead RDMA for shared data access than the compute-bound benchmarks. Even though \texttt{RDMA-WB-C-HMG} uses RDMA, this configuration brings the cache blocks from a remote GPU in its L2\$ instead of its L1\$ as in the case of \texttt{RDMA-WB-NC}. Hence, workloads that exploit temporal and spatial locality (i.e. \texttt{mm} and \texttt{conv}) achieve speed-up up to 18$\times$ for \texttt{RDMA-WB-C-HMG} configuration.

If we compare the speed-up of \texttt{SM-WB-NC} and \texttt{SM-WT-NC}, for all the compute-bound benchmarks, the difference between a WB L2\$ and a WT L2\$ is less than 1\%. But for the memory-bound benchmarks, we observe up to 3$\times$ better performance when employing a WT cache. This lower performance of WB cache can be explained by inspecting the L1\$ and L2\$ transactions. Figures~\ref{fig:L2-MM} and~\ref{fig:L1-L2} show the normalized\footnote{We use normalized values here due to the wide variations in the number of L2\$ to MM as well as L1\$ to L2\$, transactions across the different benchmarks.} L2\$ and L1\$ traffic \ignore{\footnote{We use the number of transactions to represent the amount of traffic. The total amount of traffic can be easily calculated by using a simple formula: Request traffic in bytes $=$ number of requests$\times$ number of bytes per request and Response traffic in bytes $=$ number of responses $\times$ number of bytes per response.}} in terms of number of L2\$ to MM and L1\$ to L2\$ transactions and responses, respectively, for both read and write operations. As we observe in Figures~\ref{fig:L2-MM}, as expected, when using WB there are, on average, 22.7\% less transactions than WT from L2\$ to MM for all the benchmarks. However, it is counter-intuitive that even with fewer L2\$ to MM transactions, \texttt{SM-WB-NC} performs worse than \texttt{SM-WT-NC}. For a read or write miss in the L2\$ with a WB policy, first, the L2\$ performs a write to MM to generate a cache eviction if there is either a conflict or capacity miss. Only then the L2\$ can service the pending read or write transactions. The L2\$ generating the WB becomes a bottleneck when there are frequent cache evictions. Note that the benchmarks in our evaluation use large memory footprints to generate frequent capacity and conflict misses in the L2\$. Additionally, the benchmarks are streaming in nature.  Hence, the benchmarks have frequent cache evictions, which perform worse with WB than with WT. With a WT L2\$, we do not need to write the data to the MM in case of an eviction as the updated copy of the data is always available in the MM. The transactions from L1\$ to L2\$ remain the same for both \texttt{SM-WB-NC} and \texttt{SM-WT-NC} across all benchmarks.



Figure~\ref{fig:reg-speedup} also shows that our proposed \texttt{SM-WT-C-\HALCONE} suffers, on average, a 1\% performance degradation as compared to the \texttt{SM-WT-NC} configuration. This slight performance degradation is due to more L1\$ transactions being generated for \texttt{SM-WT-C-\HALCONE} as compared to \texttt{SM-WT-NC}, as seen in Figure~\ref{fig:L1-L2}. As explained earlier, the standard benchmarks do not require any support for coherency and due to their streaming nature (which means these benchmarks continuously read and write to different cache blocks) they suffer capacity and conflict misses instead of coherency misses. For more details on this, refer to Section~\ref{sec:xtreme}. We conclude that our \HALCONE protocol is efficient as it causes, on average, a 1\% performance degradation for standard MGPU benchmarks when compared to an MGPU system with \texttt{SM-WT-NC} configuration. Moreover, compared to \texttt{RDMA-WB-NC}, an MGPU system with the \texttt{SM-WT-C-\HALCONE} configuration has, on average, 4.6$\times$ better performance. Besides, the \texttt{SM-WT-C-\HALCONE}configuration, on average, outperforms \texttt{RDMA-WB-C-HMG} configuration by 3$\times$. 

\vspace{-5pt}
\subsection{Scalability Study}
\vspace{-2pt}
We use strong scaling to explore the scalability of the MGPU-SM system with \HALCONE protocol, by varying both the GPU count and CU count while keeping the size of the workloads constant. \ignore{We use strong scaling for this study by keeping the workload size constant while varying the \#GPUs for GPU scalability study and \#CUs for CU scalability study. }

\vspace{-5pt}
\subsubsection{GPU Count Scalability Study}
For this study, we use  32 CUs per GPU as the baseline comparison point. Figure~\ref{fig:sgpu} shows the speed-up for GPU counts of 1, 2, 4, 8 and 16. Here, runtimes are normalized to that of a single GPU. On average, we achieve a 1.76$\times$, 2.74$\times$, 4.05$\times$, and 5.43$\times$ speed-up in comparison to a single coherent GPU for 2, 4, 8, and 16 GPUs, respectively. Some of the workloads (i.e., \textbf{atax}, \textbf{bicg}, \textbf{mp} and \textbf{relu}), do not scale well beyond 4 GPUs due to lesser computations available for each GPU and so do not benefit from a larger GPU count. Nonetheless, the comparison shown in Figure~\ref{fig:sgpu} confirms that our proposed \HALCONE protocol is scalable and does not limit the scalability of an MGPU-SM system.

\ignore{	
\begin{subfigure}[t]
		\centering
    \includegraphics[width=\columnwidth]{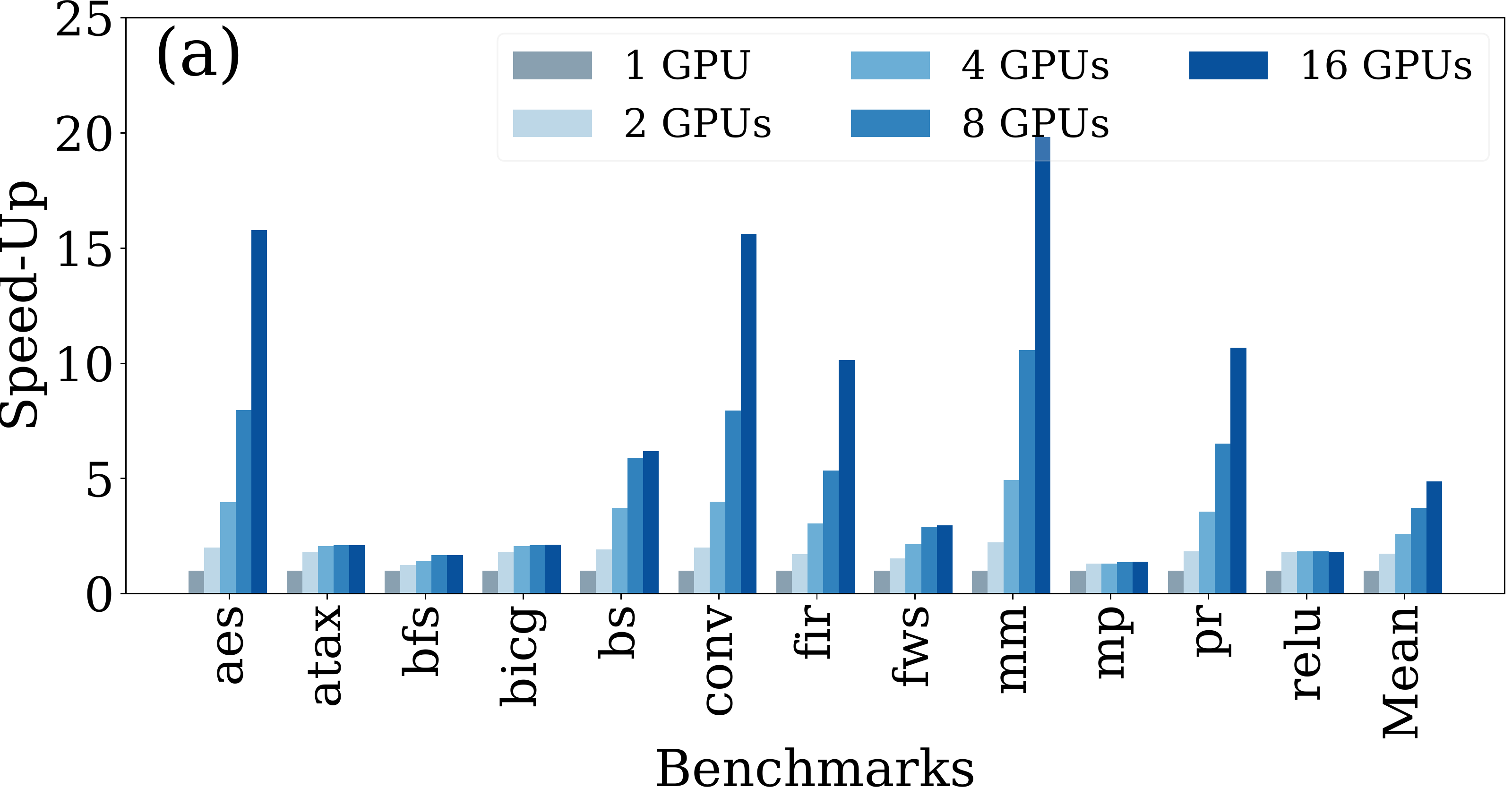}
    \caption{}\label{fig:sgpu}
	\end{subfigure}
	\begin{subfigure}[t]
    \centering
    \includegraphics[width=\columnwidth]{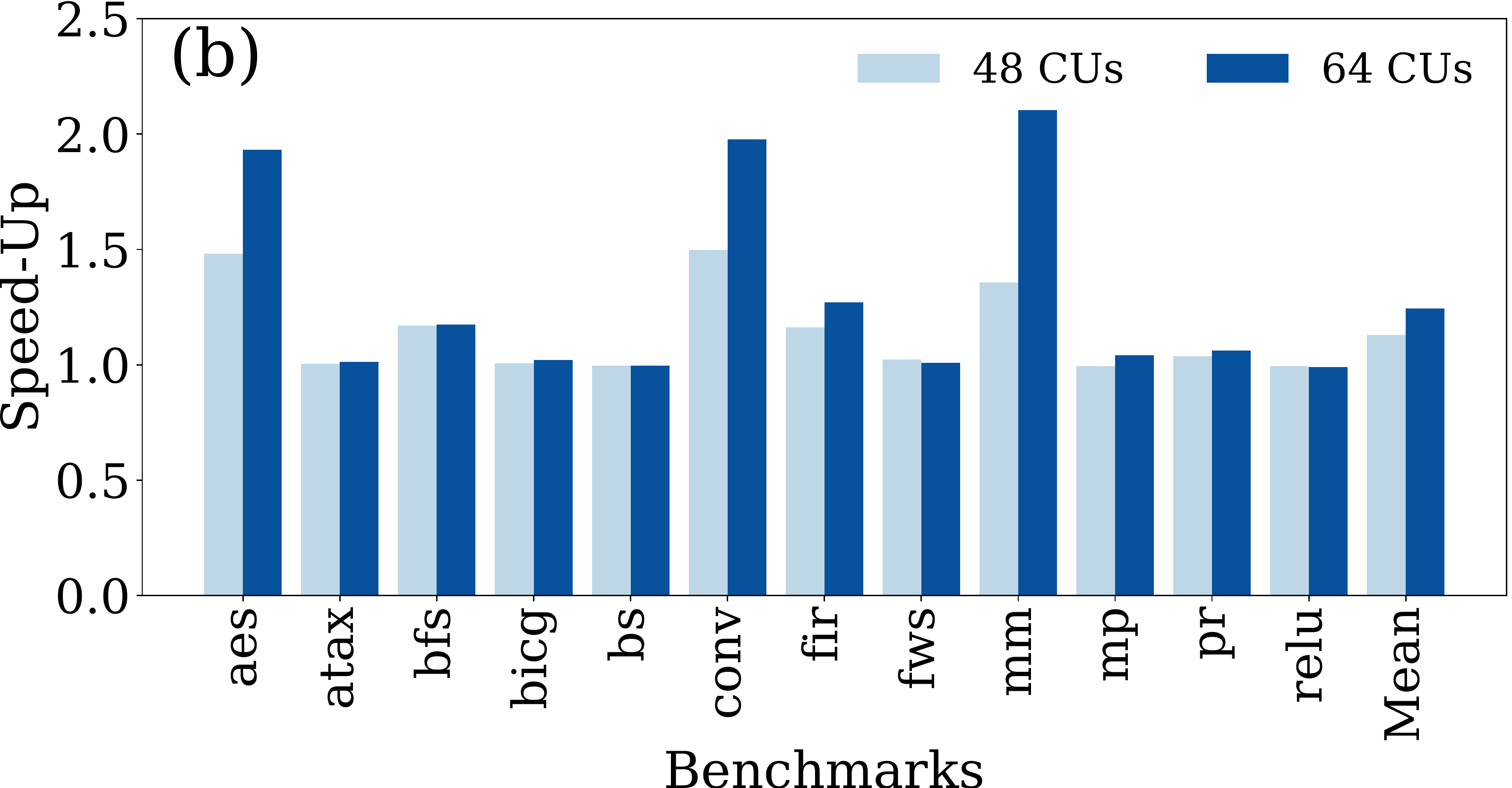}
    \caption{}\label{fig:scu-time}
	\end{subfigure}
  \begin{subfigure}[t]
    \centering
    \includegraphics[width=\columnwidth]{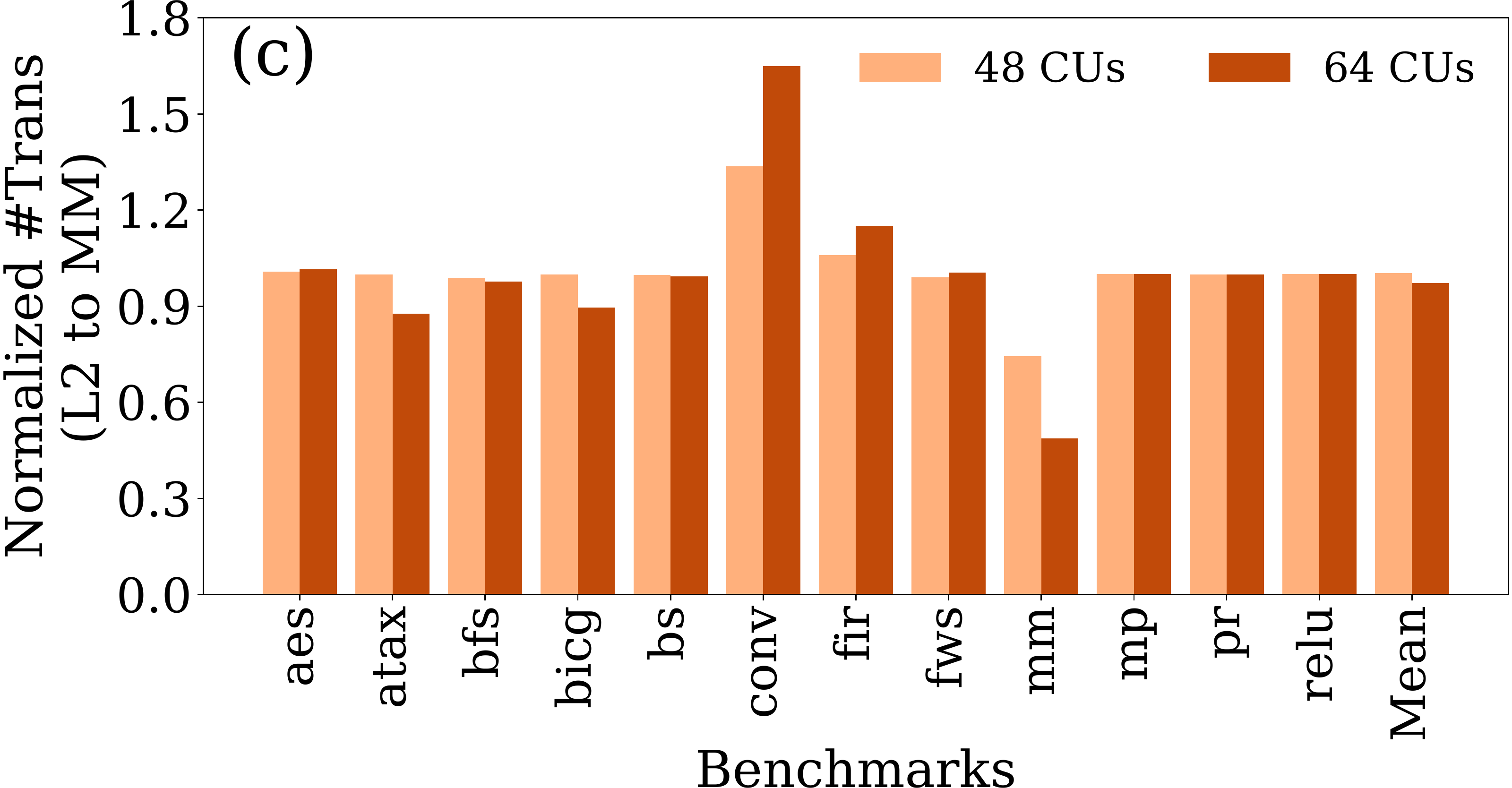}
		\caption{}\label{fig:scu-L2-MM}
	\end{subfigure}
	\vspace{-0.1in}
}

\begin{figure*} [htbp]
	\centering
	\subfigure{\label{fig:sgpu}\includegraphics[width=0.325\textwidth]{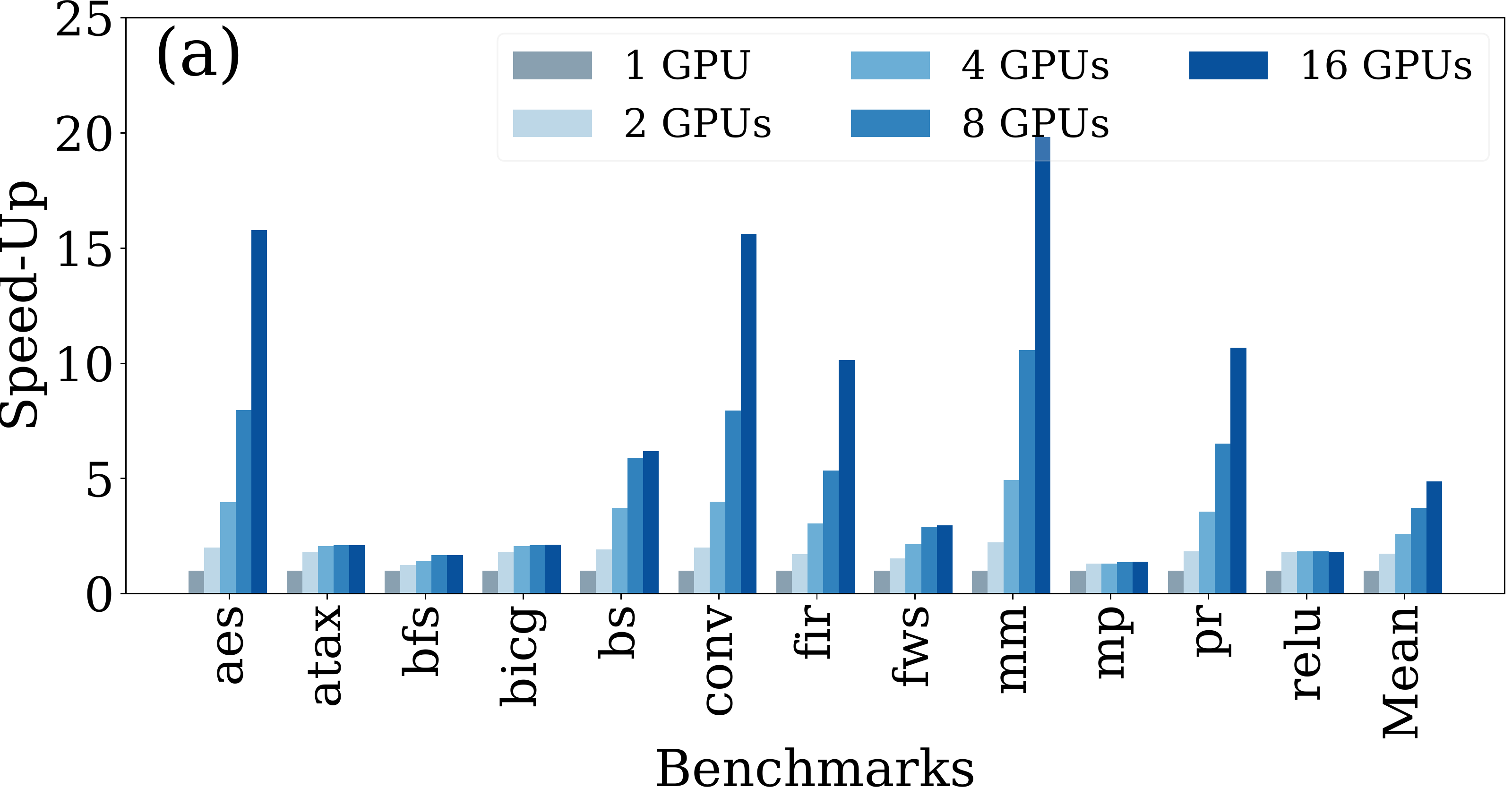}}
	\subfigure{\label{fig:scu-time}\includegraphics[width=0.325\textwidth]{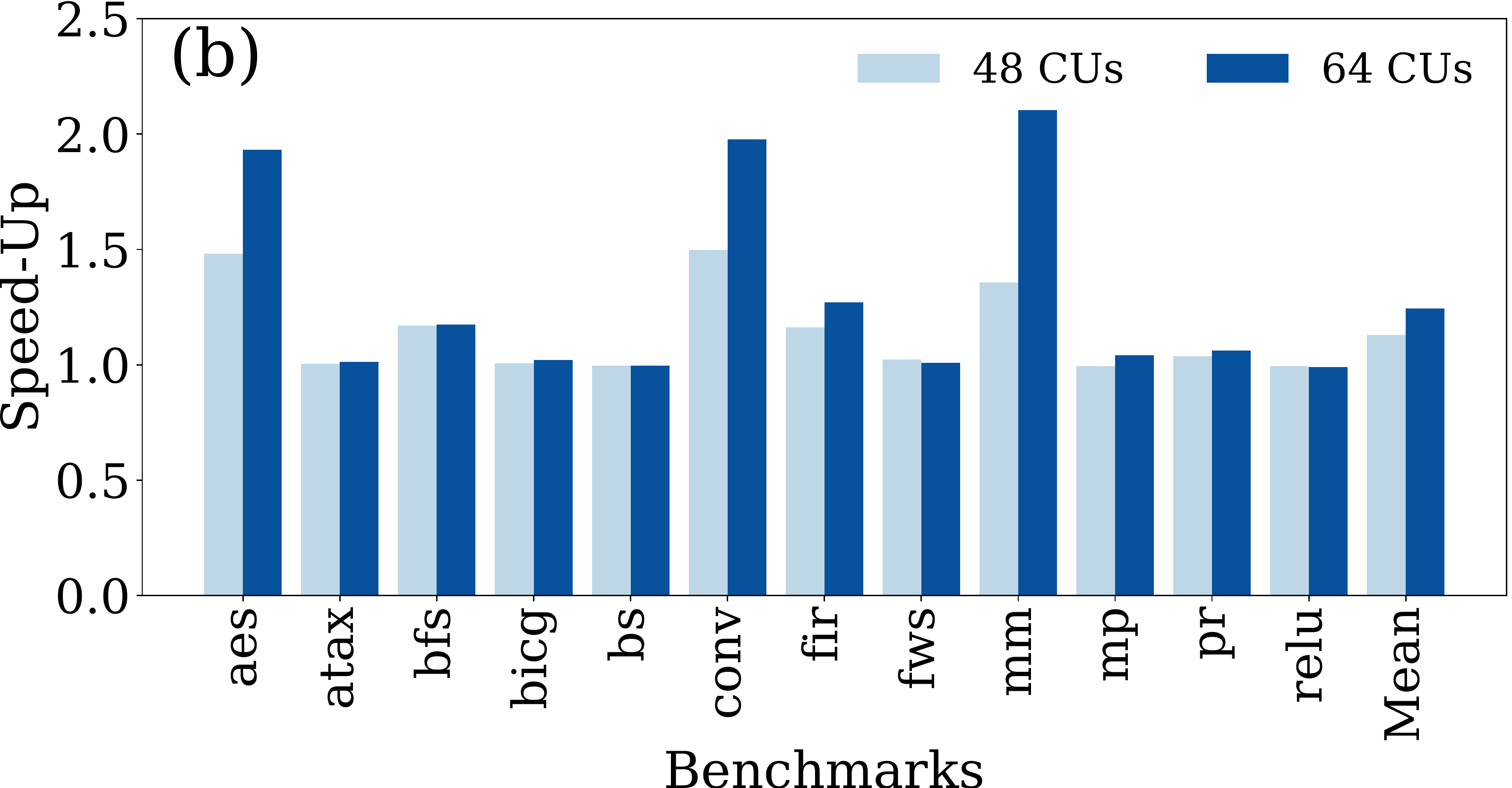}}
	\subfigure{\label{fig:scu-L2-MM}\includegraphics[width=0.325\textwidth]{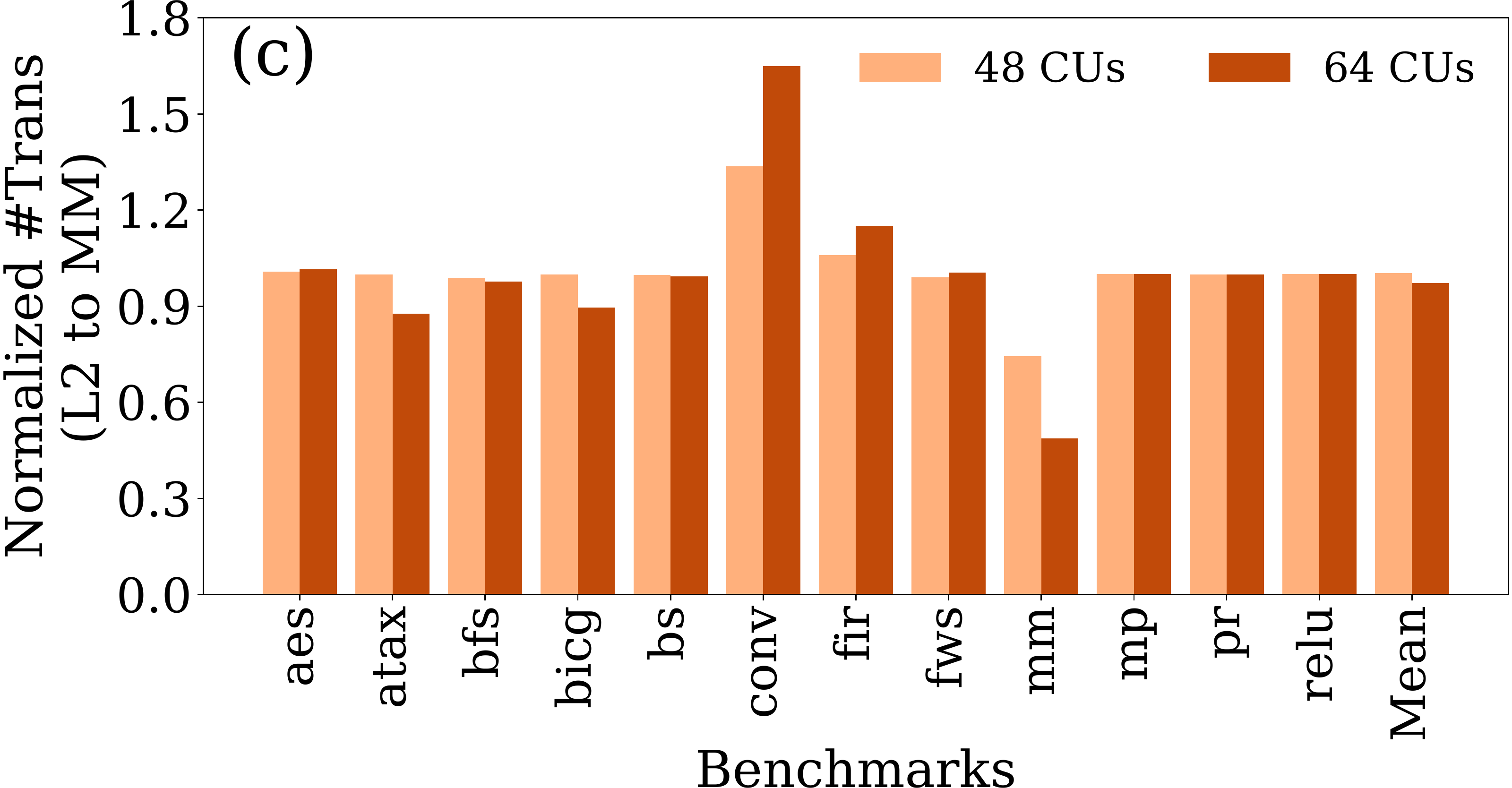}}\vspace{-8pt}
	\caption{(a) GPU scalability: Speed-up for the \texttt{SM-WT-C-\HALCONE}with different \#GPUs normalized to that of a single coherent GPU. (b) and (c) CU scalability: Speed-up and \#L2\$ transactions for the \texttt{SM-WT-C-\HALCONE}with different \#CUs normalized to that of the \texttt{SM-WT-C-\HALCONE}with 32 CUs. Mean stands for geometric mean.}
  \label{fig:tsm-scc}
	\vspace{-0.15in}
\end{figure*}


\vspace{-4pt}
\subsubsection{CU Count Scalability Study}
For this study, we use a 4-GPU system and consider 32, 48 and 64 CUs per GPU (see Figure~\ref{fig:scu-time} and Figure~\ref{fig:scu-L2-MM}). The \textbf{atax}, \textbf{bicg}, \textbf{mp}, and \textbf{relu} benchmarks do not scale with CU count as they do not have sufficient compute intensity. The \textbf{bfs} and \textbf{bs} benchmarks do not scale as we increase the CU count because of a L2\$ bottleneck. For these benchmarks, the number of transactions from L2\$ to MM for 32 CUs is the same as the number of transactions for 48 and 64 CUs. Hence, L2\$ queuing and serialization latencies dominate the runtime, irrespective of the number of CUs. In Figure~\ref{fig:L2-MM} and Figure~\ref{fig:L1-L2}, we have already demonstrated that our \HALCONE protocol, on average, introduces only 1\% additional traffic from L2\$ to MM and from L1\$ to L2\$. The \textbf{bfs} and \textbf{bs} benchmarks suffer from the L2\$ bottleneck, even when the MGPU system lacks coherency. Hence, the \HALCONE protocol itself is not a bottleneck in terms of CU scalability. The \textbf{aes}, \textbf{fir}, \textbf{mm} and \textbf{conv} benchmarks, do not stress the L2\$ even if the number of transactions from L2\$ to MM increases with the increased CU count and have sufficient compute intensity to take advantage of higher CU count. Hence, these benchmarks benefit from a larger number of CUs. On an average we see 1.12$\times$ and 1.24$\times$ speed-up as we increase the CU count from 32 to 48 and 64, respectively.



\vspace{-5pt}
\subsection{Xtreme Benchmarks} \label{sec:xtreme}
As discussed before, the standard MGPU benchmarks have been developed in accordance with the assumption that there is no hardware support for coherency and no weak-consistency programming model assumed for the GPUs. Hence, we use our synthetic benchmark suite, \textbf{Xtreme}, to evaluate the impact of our proposed \HALCONE protocol for some of the extreme cases of applications, where we need coherency to ensure the correctness of the computation. With Xtreme benchmarks, we evaluate three different scenarios:

\vspace{-0.075in}
\begin{enumerate}
  \item The data size is small, so there are neither L1\$ nor L2\$ capacity or conflict misses.
  \vspace{-0.075in}
  \item The data size is large enough to cause L1\$ capacity and conflict misses, but not large enough to cause L2\$ capacity or conflict misses.
  \vspace{-0.075in}
  \item The data size is large enough to cause both L1\$ and L2\$ capacity and conflict misses.
\end{enumerate}
\vspace{-0.075in}

\begin{figure}[t]
    \centering
    \includegraphics[width=\columnwidth]{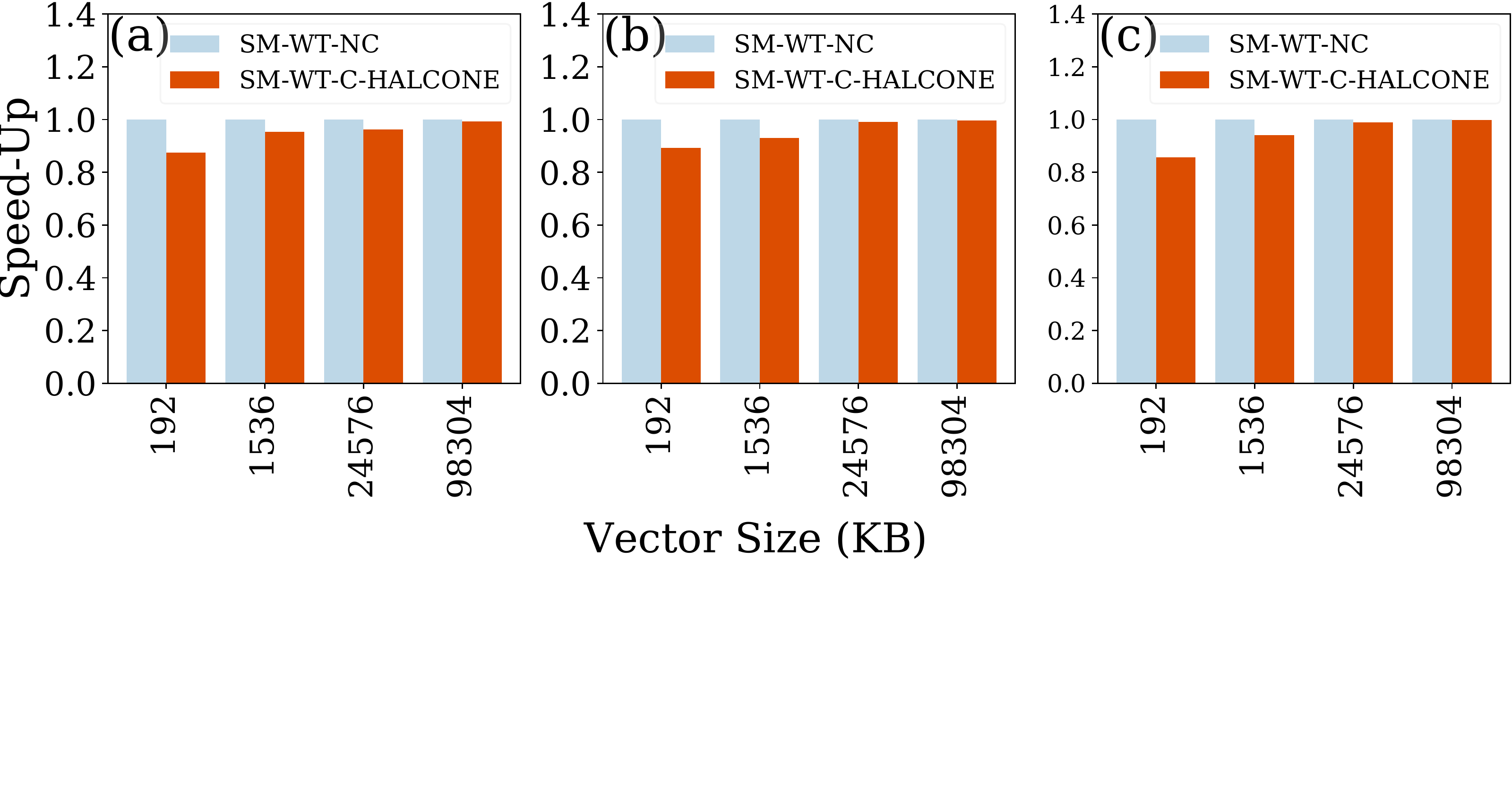}
    \vspace{-58pt}
    \caption{Speed-up for the Xtreme benchmarks running on an MGPU system with \texttt{SM-WT-C-\HALCONE} w.r.t. an MGPU system with \texttt{SM-WT-NC} for different vector sizes: (a) \textbf{Xtreme1}, (b) \textbf{Xtreme2}  and (c) \textbf{Xtreme3}}\label{fig:xtreme}
		\vspace{-0.21in}
\end{figure}

We use MGPU-SM with 4 GPUs for this evaluation. Figure~\ref{fig:xtreme} shows the comparison of speed-up for \texttt{SM-WT-NC} and \texttt{SM-WT-C-\HALCONE} for all three \textbf{Xtreme} benchmarks. The repeated writes to the same cache location in \textbf{Xtreme1} cause the \texttt{cts} of both the L1\$s and L2\$s to step ahead in logical time, leading to coherency misses for the data that was read before. For a vector size of 192 KB, we observe a performance degradation of 14.3\% for \texttt{SM-WT-C-\HALCONE}. As the vector size increases, there are more capacity and conflict misses, and eventually capacity and conflict misses far outnumber coherency misses. The coherency misses can occur if the lease expires for a cache block. However, if there are frequent cache evictions because of conflict or capacity misses, the cache blocks are evicted based on an LRU policy, even if the lease is valid. Thus, for a vector size of 98304 KB, we observe only a 0.6\% performance degradation for \texttt{SM-WT-C-\HALCONE} in comparison to \texttt{SM-WT-NC}. The \textbf{Xtreme2} benchmark exploits intra-GPU coherency. \textbf{Xtreme3} requires inter-GPU coherency among the MGPUs for correctness. The data dependency in \textbf{Xtreme2} and \textbf{Xtreme3} results in coherency misses when repeated writes are performed. We observe a performance degradation of up to 12.1\% and 16.8\% for \textbf{Xtreme2} and \textbf{Xtreme3}, respectively. This degradation decreases as the data size increases due to the corresponding increase in 
capacity and conflict misses in L1\$s and L2\$s.

\vspace{-5pt}
\subsection{Sensitivity to Timestamps}
\label{sec:ts}
We used (\texttt{RdLease}, \texttt{WrLease}) = (5, 10) for our evaluations. We examined the impact of using different (\texttt{RdLease}, \texttt{WrLease}) values of (2, 10), (10, 2), (5, 10), (10, 5), (20, 10), and (10, 20) using the coherency-aware \textbf{Xtreme} benchmarks. We found that if the difference between the \texttt{WrLease} and the \texttt{RdLease} is increased to 10 from 5, then the benchmark performance degrades by up to 3\% for the \textbf{Xtreme}. So we need to maintain a smaller difference between \texttt{RdLease} and \texttt{WrLease}. In terms of absolute values of \texttt{RdLease} and \texttt{WrLease}, a large value of the \texttt{RdLease} can help an application that performs significantly smaller number of writes than number of reads. On the other hand, a smaller \texttt{RdLease} results in more coherency misses. We choose a smaller \texttt{WrLease} value than \texttt{RdLease} value based on the assumption that if a CU or a GPU writes to a cache block, it may write to the same cache block in the future. This, in turn, prevents consecutive writes to the same block and avoids making \texttt{cts} too large, potentially causing many coherency misses.

\vspace{-6pt}
\section{Related Work}\label{sec:related}
MGPU systems have been recently adopted as the computing platform of choice by variety of data-intensive applications. Today there is growing interest in developing optimized MGPU system architectures and efficient hardware coherence support to reduce programming complexity.

\textbf{MGPU System Design:}
Milic et al. propose a NUMA-aware multi-socket GPU solution to resolve performance bottlenecks related to NUMA memory placement in multi-socket GPUs~\cite{milic2017beyond}. The proposed system dynamically adapts inter-socket link bandwidth and caching policies to avoid NUMA effects. 
\ignore{The proposed system exploits the changing application phase behavior in terms of inter-socket bandwidth demand and data locality in L1 and L2 caches, by dynamically adapting the inter-socket link bandwidth and L1 and L2 caching policies, respectively.}
Our CC-MGPU system completely avoids the impact of NUMA on performance.
Arunkumar et al.~\cite{arunkumar2017mcm} and  Ren et al.~\cite{hmg} propose a MCM-GPU, where multiple GPU modules are integrated in a package to improve energy efficiency. As in MCM-GPU, our CC-MGPU can take advantage of novel integration technologies to improve energy efficiency and performance. Arunkumar et al.~\cite{arunkumar2019understanding} also argue the need to improve inter-GPU communication. We plan to explore high-bandwidth network architectures for CC-MGPU systems in the future.  
\ignore{We will consider this as future work.}

\textbf{MGPU Coherency:}
NUMA-Aware multi-socket GPU \cite{milic2017beyond} maintains inter-GPU coherency by extending SW-based coherence for L1\$s to the L2\$s. The resulting coherency traffic lowers application performance. Similarly, MCM-GPU~\cite{arunkumar2017mcm} leverages the software-based L1\$ coherence protocol for its L1.5\$. The flushing of the caches and coherency traffic 
\ignore{Apart from expensive flushing of both caches that are forced at every synchronization event, non-negligible coherence-related invalidation traffic}
hurt system scalability. Young et al.~\cite{young2018combining} propose CARVE method, where part of a GPU's memory is used as a cache for shared remote data and the GPU-VI protocol is used for coherency. This protocol does not scale well with an increase in the amount of read-write transactions and false sharing. Also, the CARVE method can cause performance degradation for workloads with large memory footprint as it reduces effective GPU memory space. To reduce coherency traffic, Singh et al. propose timestamp-based coherency (TC) protocol for intra-GPU
coherency~\cite{singh2013cache}. As this protocol relies on a globally
synchronized clock across all CUs, maintaining clock synchronization is a
challenging task for large MGPU systems. To address this, Tabakh et
al.~\cite{tabbakh2018g} propose a logical timestamp based coherence protocol
(G-TSC). However, as discussed in Section~\ref{sec:G-TSC}, the G-TSC protocol is designed for single GPU systems and does not scale well for MGPU systems.
\ignore{
Due to
the promising results achieved by G-TSC, we extend and modify the G-TSC protocol
for the CC-MGPU systems. Power et al.~\cite{power2013heterogeneous} propose
different cache coherence protocol for APUs. In contrast, we are the first to
propose an efficient cache coherence protocol for a CC-MGPU that relies on existing weak consistency models. Sinclair et al.\mbox{~\cite{sinclair2015efficient}} propose a memory consistency model for GPUs without using user-managed scopes of a variable in the memory hierarchy, while Alsop et al.\mbox{~\cite{kumar2015fusion}} propose a lazy release
consistency model for GPUs. Both of these studies only consider single CPU and single
GPU systems. Exploring efficient multi-GPU memory consistency models is an open
challenge and results from these studies can be leveraged to address that challenge. 
}
HMG~\cite{hmg} is a recent hardware-managed cache coherence protocol for distributed L2\$s in MCM-GPUs using a scoped memory model consistency. HMG proposes to extend a simple VI-like protocol to track sharers in a hierarchical way that is tailored to the MCM-GPU architecture, and achieves a cost-effective solution in terms of on-chip area overhead, inter-GPU coherence traffic reduction and high performance.
This protocol, however, relies on error-prone scoped memory consistency model that increases programming complexity. In contrast, our new timestamp-based coherence \HALCONE protocol operates assuming weak consistency model, which is currently adopted by modern GPU products, and is able to outperform HMG by 3$\times$ on average (see Section~\ref{sec:regular}).

\vspace{-5pt}
\section{Conclusion}
In this work, we propose a novel MGPU system, where GPUs physically share the MM. This system eliminates the programmer's burden of unnecessary data replication and expensive remote memory accesses. To ensure seamless sharing of data across and within multiple GPUs, we propose \HALCONE, a novel timestamp-based coherence protocol. For standard benchmarks, a MGPU-SM system (that has 4 GPUs and uses \HALCONE) performs on average, 4.6$\times$ faster than the non-coherent conventional MGPU system with same number of GPUs. In addition, compared to a coherent MGPU system using the state-of-the-art HMG coherence protocol, an MGPU system that uses \HALCONE reports 3$\times$ higher performance. Our scalability study shows that our coherence protocol scales well in terms of both GPU count and CU count. We develop synthetic benchmarks that leverage data sharing to examine the impact of our \HALCONE protocol on performance. For the worst case scenario in our synthetic benchmarks, the proposed MGPU-SM with \HALCONE suffers from only a 16.8\% performance overhead.
\ignore{
\section{List of Actions}

\subsection{What has been done so far}

\begin{itemize}
\item Initial thermal simulation
\item Implementation of unoptimized time stamp based coherence protocol
\item Three working configurations:
\begin{enumerate}
\item TSM, L1 Write-through, L2 Write-through, No Coherency
\item TSM, L1 Write-through, L2 Write-back, No Coherency
\item TSM, L1 Write-through, L2 Write-through, With Coherency
\end{enumerate}
\end{itemize}

\subsection{To be done}
\begin{enumerate}
\item Optimization of the timestamp based protocol
\item Implementation of baseline with RDMA
\item Updating synthetic benchmarks
\item Implementation of coherency policy with WB
\item Improved placement for reducing temperature in the hotspot
\item Run Experiments with  2, 4, 8 and 16 GPUs
\end{enumerate}

 \subsection{Needs to be discussed}
 \begin{itemize}
 \item Implementation of RDMA and baseline
 \item What else need to be in the paper
 \end{itemize}
}


\bibliographystyle{IEEEtranS}
\bibliography{tsm1,kaeli,trinayan,ajay}
\balance
\newpage


\end{document}